\documentclass[fleqn,usenatbib]{mnras}

\usepackage{newtxtext,newtxmath}

\usepackage[T1]{fontenc}

\DeclareRobustCommand{\VAN}[3]{#2}
\let\VANthebibliography\thebibliography
\def\thebibliography{\DeclareRobustCommand{\VAN}[3]{##3}\VANthebibliography}

\usepackage{graphicx}	
\usepackage{amsmath}	
\usepackage{multirow} 
\usepackage{natbib}
\usepackage{tabu} 
\usepackage{orcidlink}

\title[Red QSOs in DESI]{A striking relationship between dust extinction and radio detection in DESI QSOs: evidence for a dusty blow-out phase in red QSOs}

\author[V. A. Fawcett et al.]
{V. A. Fawcett,$^{1,2}$\thanks{E-mail: vicky.fawcett@newcastle.ac.uk}\orcidlink{0000-0003-1251-532X}
D. M. Alexander,$^{2}$\orcidlink{0000-0002-5896-6313}
A. Brodzeller,$^{3}$
A. C. Edge,$^{2}$
D. J. Rosario,$^{1,2}$ 
A. D. Myers,$^{4}$ 
\newauthor
J.~Aguilar,$^{5}$
S.~Ahlen,$^{6}$
R. Alfarsy,$^{7}$ 
D.~Brooks,$^{8}$
R. Canning,$^{7}$ 
C. Circosta,$^{8}$ 
K.~Dawson,$^{3}$
A.~de la Macorra,$^{9}$\newauthor
P.~Doel,$^{8}$
K.~Fanning,$^{10}$
A.~Font-Ribera,$^{11}$
J.~E.~Forero-Romero,$^{12}$
S.~Gontcho A Gontcho,$^{5}$
J.~Guy,$^{5}$\newauthor
C. M. Harrison,$^{1}$
K.~Honscheid,$^{10,13,14}$
S.~Juneau,$^{15}$ 
R.~Kehoe,$^{16}$
T.~Kisner,$^{5}$ 
A.~Kremin,$^{5}$
M.~Landriau,$^{5}$
\newauthor
M.~Manera,$^{11}$
A. M. Meisner,$^{15}$
R.~Miquel,$^{11,17}$
J.~Moustakas,$^{18}$
J.~Nie,$^{19}$
W.~J.~Percival,$^{20,21,22}$
C.~Poppett,$^{5,23,24}$\newauthor
R.~Pucha,$^{25}$
G.~Rossi,$^{26}$
D.~Schlegel,$^{5}$
M. Siudek,$^{11,27}$
G.~Tarl\'{e},$^{28}$ 
B.~A.~Weaver,$^{15}$
Z.~Zhou,$^{19}$
H.~Zou$^{19}$
\\
$^{1}$School of Mathematics, Statistics and Physics, Newcastle University, NE1 7RU, UK\\
$^{2}$Centre for Extragalactic Astronomy, Department of Physics, Durham University, South Road, Durham, DH1 3LE, UK\\
$^{3}$Department of Physics and Astronomy, The University of Utah, 115 South 1400 East, Salt Lake City, UT 84112, USA\\
$^{4}$Department of Physics \& Astronomy, University  of Wyoming, 1000 E. University, Dept.~3905, Laramie, WY 82071, USA\\
$^{5}$Lawrence Berkeley National Laboratory, 1 Cyclotron Road, Berkeley, CA 94720, USA\\
$^{6}$Physics Dept., Boston University, 590 Commonwealth Avenue, Boston, MA 02215, USA\\
$^{7}$Institute of Cosmology \& Gravitation, University of Portsmouth, Dennis Sciama Building, Portsmouth, PO1 3FX, UK\\
$^{8}$Department of Physics \& Astronomy, University College London, Gower Street, London, WC1E 6BT, UK\\
$^{9}$Instituto de F\'{\i}sica, Universidad Nacional Aut\'{o}noma de M\'{e}xico,  Cd. de M\'{e}xico  C.P. 04510,  M\'{e}xico\\
$^{10}$The Ohio State University, Columbus, 43210 OH, USA\\
$^{11}$Institut de F\'{i}sica d’Altes Energies (IFAE), The Barcelona Institute of Science and Technology, Campus UAB, 08193 Bellaterra Barcelona, Spain\\
$^{12}$Departamento de F\'isica, Universidad de los Andes, Cra. 1 No. 18A-10, Edificio Ip, CP 111711, Bogot\'a, Colombia\\
$^{13}$Center for Cosmology and AstroParticle Physics, The Ohio State University, 191 West Woodruff Avenue, Columbus, OH 43210, USA\\
$^{14}$Department of Physics, The Ohio State University, 191 West Woodruff Avenue, Columbus, OH 43210, USA\\
$^{15}$NSF's NOIRLab, 950 N. Cherry Ave., Tucson, AZ 85719, USA\\
$^{16}$Department of Physics, Southern Methodist University, 3215 Daniel Avenue, Dallas, TX 75275, USA\\
$^{17}$Instituci\'{o} Catalana de Recerca i Estudis Avan\c{c}ats, Passeig de Llu\'{\i}s Companys, 23, 08010 Barcelona, Spain\\
$^{18}$Department of Physics and Astronomy, Siena College, 515 Loudon Road, Loudonville, NY 12211, USA\\
$^{19}$National Astronomical Observatories, Chinese Academy of Sciences, A20 Datun Rd., Chaoyang District, Beijing, 100012, P.R. China\\
$^{20}$Department of Physics and Astronomy, University of Waterloo, 200 University Ave W, Waterloo, ON N2L 3G1, Canada\\
$^{21}$Perimeter Institute for Theoretical Physics, 31 Caroline St. North, Waterloo, ON N2L 2Y5, Canada\\
$^{22}$Waterloo Centre for Astrophysics, University of Waterloo, 200 University Ave W, Waterloo, ON N2L 3G1, Canada\\
$^{23}$Space Sciences Laboratory, University of California, Berkeley, 7 Gauss Way, Berkeley, CA  94720, USA\\
$^{24}$University of California, Berkeley, 110 Sproul Hall \#5800 Berkeley, CA 94720, USA\\
$^{25}$Steward Observatory, University of Arizona, 933 N, Cherry Ave, Tucson, AZ 85721, USA\\
$^{26}$Department of Physics and Astronomy, Sejong University, Seoul, 143-747, Korea\\
$^{27}$Institute of Space Sciences, ICE-CSIC, Campus UAB, Carrer de Can Magrans s/n, 08913 Bellaterra, Barcelona, Spain\\
$^{28}$University of Michigan, Ann Arbor, MI 48109, USA\\
}

\date{Accepted XXX. Received YYY; in original form ZZZ}

\pubyear{2023}

\begin{document}
\defcitealias{hamann}{H17}
\defcitealias{glikman22}{G22}
\defcitealias{fawcett22}{F22}

\label{firstpage}
\pagerange{\pageref{firstpage}--\pageref{lastpage}}
\maketitle

\begin{abstract}
We present the first eight months of data from our secondary target program within the ongoing Dark Energy Spectroscopic Instrument (DESI) survey. Our program uses a mid-infrared and optical colour selection to preferentially target dust-reddened QSOs that would have otherwise been missed by the nominal DESI QSO selection. So far we have obtained optical spectra for 3038 candidates, of which $\sim$\,$70$ per~cent of the high-quality objects (those with robust redshifts) are visually confirmed to be Type~1 QSOs, consistent with the expected fraction from the main DESI QSO survey. By fitting a dust-reddened blue QSO composite to the QSO spectra, we find they are well-fitted by a normal QSO with up to $A_V$\,$\sim$\,$4$\,mag of line-of-sight dust extinction. Utilizing radio data from the LOFAR Two-metre Sky Survey (LoTSS) DR2, we identify a striking positive relationship between the amount of line-of-sight dust extinction towards a QSO and the radio detection fraction, that is not driven by radio-loud systems, redshift and/or luminosity effects. This demonstrates an intrinsic connection between dust reddening and the production of radio emission in QSOs, whereby the radio emission is most likely due to low-powered jets or winds/outflows causing shocks in a dusty environment. On the basis of this evidence we suggest that red QSOs may represent a transitional ``blow-out'' phase in the evolution of QSOs, where winds and outflows evacuate the dust and gas to reveal an unobscured blue QSO.
\end{abstract}


\section{Introduction}

Quasi-stellar objects (QSOs), also known as quasars, are the most powerful class of Active Galactic Nuclei (AGN), with extremely high bolometric luminosities ($10^{45}$--$10^{48}$\,erg\,s$^{-1}$). These high luminosities are now known to be due to mass accretion onto a supermassive black-hole (SMBH; masses of $10^7$--$10^9$\,M$_{\odot}$).
The observed correlation between the masses of SMBHs at the centers of spheroidal galaxies and the bulge mass of the hosting galaxy suggests an intrinsic link between the growth of the SMBH and the surrounding host galaxy \citep{kormendy}.

A canonical model for this growth phase invokes gas-rich major mergers \citep{sanders,hop,alex_hick_12} in which rapid gas inflow to the nucleus fuels a powerful QSO, obscured in its early stages by a high column of gas and dust. Through powerful winds and/or outflows, an unobscured ``typical'' blue QSO is eventually revealed. However, it is still unclear what type of object represents the transitional ``blow-out'' stage between the obscured and unobscured QSO phase. Potential candidates include; 1) the well-studied dust obscured ``red QSOs'' (e.g., \citealt{white03,Glikman_2004,glik7,maddox,klindt}), selected as broad line QSOs with red optical or near-infrared (NIR) colours; 2) ``Broad Absorption Line Quasars'' (BALQSOs; e.g., \citealt{BI,morabito,petley}), which display broad absorption typically blueward of the C\,{\sc iv} emission line; and 3) the rarer ``Extremely Red Quasars'' (ERQs; e.g., \citealt{ross,hamann}), selected to have red optical--mid-infrared (MIR) colours. All of these objects are thought to play an important role in the regulation of star formation in the host galaxy (otherwise known as ``AGN feedback''; see reviews by \citealt{heckman} and \citealt{harrison_2018}). 

Over recent years, a number of groups have explored whether red QSOs represent a transitional blow-out phase, determining their properties and how they relate to typical blue QSOs \citep{richards,Glikman_2004,glik7,glik12,glikman22,georg,Urrutia_2009,ban12,ban,kim18,klindt,rosario,rosario_21,fawcett20,fawcett21,fawcett22,calistro,petter,stacey}. In our recent work, we have shown that the majority of red QSOs from the Sloan Digital Sky Survey (SDSS; \citealt{sdss_instrument}), selected based on their optical $g-i$ colour, are red due to line-of-sight dust extinction (\citealt{calistro,fawcett22}). Somewhat surprisingly we found that red QSOs exhibited enhanced radio emission, compared to blue QSOs, which is typically compact ($<$\,$2$\,kpc; \citealt{klindt,fawcett20,rosario,rosario_21}) and peaks around the radio-quiet/radio-loud threshold \citep{fawcett20,rosario}. The observed differences in the radio properties of red and blue QSOs do not appear to be associated with any differences in the star formation (\citealt{fawcett20,calistro,andonie}), accretion \citep{klindt,fawcett22}, or large scale structure \citep{petter} properties between the two populations. On the basis of these results, a potential self-consistent scenario that links the enhanced radio emission to the dust in red QSOs is that the radio emission is due to shocks produced by either an outflow or a jet interacting with a higher opacity interstellar medium (ISM)/circumnuclear environment. These outflows/jets could also potentially drive out the surrounding gas and dust (i.e., AGN feedback), eventually revealing a typical blue QSO; therefore, red QSOs are prime candidates for an intermediate stage in the evolution of QSOs.

It is worth noting that the measured dust extinctions towards optically selected SDSS red QSOs are modest ($A_V$\,$<$\,$0.7$\,mag; \citealt{calistro,fawcett22}) and they likely represent the tip of the iceberg of a larger, more heavily reddened QSO population. Indeed, other studies have utilized NIR photometry in order to select more heavily reddened red QSO samples (e.g., \citealt{ban,Glikman_2004,glik7,glikman22}), with measured dust extinctions up to $A_V$\,$\sim$\,$ 5$\,mag. For example, \cite{glikman22} used a \textit{WISE}--2MASS colour selection to define a sample of red QSOs with dust extinctions $E(B-V)>0.25$\,mag ($A_V\gtrsim0.8$\,mag)\footnote{$A_V$\,$=$\,$R_V$\,$\times$\,$E(B-V)$; where $R_V$ is the total-to-selective extinction, which determines the shape of the extinction curve \citep{cardelli}. For example, the Milky Way dust extinction curve has an $R_V=3.1$ \citep{fitz}.}. Utilizing radio data from both the Faint Images of the Radio Sky at Twenty-centimeters (FIRST; \citealt{becker}) and the Very Large Array Sky Survey (VLASS; \citealt{vlass}), they found a higher radio detection fraction for their red QSOs compared to a blue QSO sample ($E(B-V)<0.25$\,mag), in addition to a higher fraction of compact radio morphologies for the red QSOs in both radio datasets. These results are remarkably similar to our less reddened optically selected red QSOs \citep{klindt,fawcett20,rosario}.

A limitation of our previous red QSO studies is the shallow magnitude limit of SDSS which restricts us to the most luminous and optically brightest objects, with only modest amounts of extinction. We therefore require larger, fainter, and redder samples of QSOs to robustly test whether red QSOs represent a transitional phase in the evolution of QSOs: the Dark Energy Spectroscopic Instrument (DESI; \citealt{desi,desiII}) has the capability to do just this. DESI has completed the first year of a five year survey which aims to obtain optical spectra for $\sim$\,$3$\,million QSOs down to $r$\,$\sim$\,$23.2$\,mag \citep{desi_instrument,silber,DESIcorrector2023}. These data will be revolutionary in QSO science, pushing 1--2 mag fainter than SDSS and, consequently, identifying both more obscured and more typical systems (see Section~5 of \citealt{VI1} for diversity in the QSO spectra).

The DESI QSO target selection consists of both an optical colour selection and a Random Forest selection \citep{yeche,QSO_desi}\footnote{The code for the various Survey Validation selections are publicly available on GitHub: \url{https://github.com/desihub/desitarget/blob/0.51.0/py/desitarget/sv1/sv1_cuts.py}}. 
Like any standard optical QSO selection, both the Random Forest and colour selection are biased towards optically blue QSOs since the majority of QSOs, and therefore the Random Forest training sets, are blue. This bias will inevitably select against the more ``exotic'' and dust reddened QSO population. In October 2020, a call for DESI proposals for ``secondary target'' programs was announced, with the aim of utilizing spare DESI fibres and expanding the DESI science beyond the key cosmological aims of the survey. Our group successfully proposed a program (PI: V. Fawcett) which targets dust-reddened QSOs that have been missed by the nominal DESI QSO selections, with aims to: 1) push to higher levels of dust extinction than can be identified by either the SDSS or the nominal DESI QSO selection, 2) determine whether the enhanced radio emission found in SDSS red QSOs extends to more heavily reddened systems, 3) explore the origin of the enhanced radio emission found in red QSOs, and 4) explore exotic sub-populations of QSOs within DESI.

In this first paper we introduce and present the first eight months of data from our DESI secondary target program and explore the connection between dust extinction and radio emission in DESI QSOs, using data from the LOFAR Two-metre Sky Survey (LoTSS) Data Release 2 (\citealt{shim}). In Section~\ref{sec:method_desi_paper} we describe our secondary target program and in Section~\ref{sec:final_qso} we describe how our secondary target program extends the nominal DESI QSO selection, the methods used in this paper, and the construction of our combined QSO sample which is used for the scientific analyses in this paper. In Section~\ref{sec:results} we present our results and in Section~\ref{sec:dust_desi_rad_dis} we discuss how our sample extends previous red QSO work and explore the origin of the radio emission. Unless specified otherwise, in this paper we use AB magnitudes. Throughout our work we adopt a flat $\uplambda$-cosmology with $H_0$\,$=$\,70~km\,s$^{-1}$Mpc$^{-1}$, $\Omega$\textsubscript{M}\,$=$\,0.3~and~$\Omega_{\uplambda}$\,$=$\,0.7. The data underlying the figures in this paper are available online: \url{https://doi.org/10.5281/zenodo.8147342}.

\section{Secondary target program}\label{sec:method_desi_paper}
In this section we describe our secondary target selection and the DESI data (Section~\ref{sec:sample_desi}) and spectral classification method (Section~\ref{sec:method_VI}).

\subsection{Defining our secondary target program selection}\label{sec:sample_desi}
\begin{figure*}
    \centering
    \includegraphics[width=0.97\textwidth]{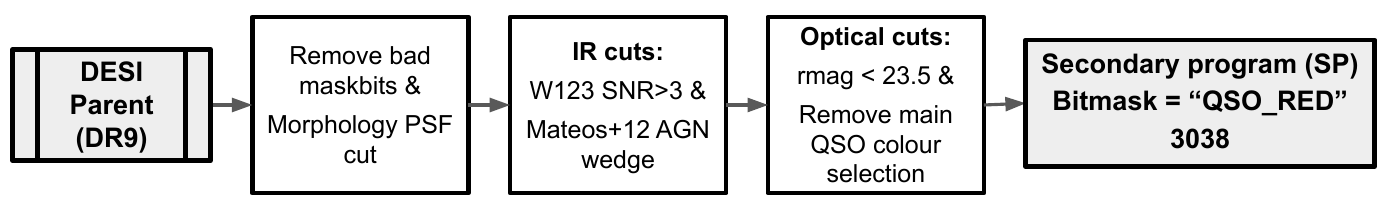}
    \caption{Flow chart illustrating the selection process for our DESI secondary program (SP) sample. The DESI SP sample consists of objects that were specifically targeted by our secondary target program; these must satisfy a morphology, magnitude, and \textit{WISE} $W123$ signal-to-noise cut, have MIR colours consistent with the \textit{WISE} \protect\cite{mateos} AGN wedge, and pass an optical colour-colour cut (see Fig.~\ref{fig:colour_colour}), resulting in 3038 objects that were observed. Each step is described in more detail in Section~\ref{sec:sample_desi}. An electronic table containing the DESI \texttt{TARGETID} for the full SP sample, and additionally the RA, Dec, redshift, spectype, and $L_{\rm 6\,\upmu m}$ for the SP sample included in the DESI Early Data Release \citep{sv}, is available online. For the analyses in this paper, we further restrict the sample to visually confirmed QSOs and combine with typical DESI QSOs, as is illustrated in Fig.~\ref{fig:flow2}.}
    \label{fig:flow}
\end{figure*}

Our secondary target program aims to expand the DESI QSO colour space, utilizing MIR data from the Wide-field Infrared Survey Explorer (\textit{WISE}; \citealt{wise}). The main goal of this program is to build up a statistically significant sample of dust-reddened QSOs that may have otherwise been missed by the nominal DESI QSO selections. However, due to our optical colour selection (which we describe in more detail below) our program will also select extremely blue QSOs. Therefore it should be noted that, although the bitmask\footnote{Different types of targets are assigned different bits, which can then be accessed via the targeting bitmasks (e.g., \texttt{DESI\_TARGET}). For example, in DESI the \texttt{QSO} targets are assigned a \texttt{desi\_mask} bit 2 and our secondary \texttt{QSO\_RED} targets are assigned a \texttt{scnd\_mask} bit 5. Therefore, the targets that are solely targeted by our program would be selected by \texttt{SCND\_TARGET}\,$=$\,$32$ ($\equiv 2^5$). For more information, see \cite{myers}.} assigned to the objects observed by our program is \texttt{QSO\_RED}, the sample will include a small fraction of optically blue QSOs, in addition to dust-reddened QSOs. For the analyses in this paper, our secondary target program is solely used to boost the source statistics of the rarer dust-reddened QSOs and in combination with a larger nominal sample of DESI QSOs, the majority of which will be typical blue QSOs by selection.

Our current sample (which we refer to as the ``DESI secondary program (SP) sample'') is taken from the first $\sim$\,$8$ months of observations, including all of the DESI Survey Validation (SV; \citealt{sv}) data (Dec 2020--June 2021) and part of the `Main' survey data (May 2021--July 2021).\footnote{The DESI Early Data Release (EDR; \citealt{sv}) will only include observations from the SV period.} The DESI SV observations are split into three different targeting campaigns: `SV1', `SV2', and `SV3'; our DESI SP sample only utilizes SV1 and SV3 observations since no secondary targets were assigned during SV2 \citep{myers,sv}. During the SV3 phase, primary targets were prioritized for multiple observations before secondary targets were scheduled in order to obtain highly complete samples of DESI primary targets. Thus, a portion of our SP sample is biased towards the colours of primary DESI classes, particularly Luminous Red Galaxies (LRGs; \citealt{lrg}; $z$\,$<$\,$21.7$ limit in SV3) and QSOs (\citealt{QSO_desi}; $r$\,$<$\,$23$ limit in SV3). This bias will gradually diminish over the course of the DESI Main survey, for which our SP targets are assigned an observational priority at least as high as any other primary target (see Section 5 of \citealt{schlafly}). The flowchart in Fig.~\ref{fig:flow} displays our secondary target selection process, which we describe below. 

A key distinguishing component of our SP compared to the DESI QSO sample is the MIR \textit{WISE} selection. We first started with the DESI DR9 legacy photometric catalogues \citep{dey}\footnote{\url{https://www.legacysurvey.org/}}, which contain optical photometry in the $g$- (4730\,\AA), $r$- (6420\,\AA), and $z$-bands (9260\,\AA), as well as four MIR bands (at 3.4, 4.6, 12, and 22\,$\upmu$m) from stacks of \textit{WISE/NEOWISE} data \citep{wise,neowise,neowise2}, referred to as unWISE coadds (\citealt{unwise}; for more details see Section~5 of \citealt{dey}). We first used the \texttt{MASKBIT} column in order to remove objects with flagged photometric errors. In our selection we remove objects with a \texttt{MASKBIT} corresponding to either \texttt{BRIGHT} (object touches the pixel of a bright star), \texttt{ALLMASK\_G}, \texttt{ALLMASK\_R}, \texttt{ALLMASK\_Z} (object touches a bad pixel in either the $g$-, $r$-, or $z$-band images), \texttt{BAILOUT} (object touches a pixel in a region where source fitting failed), \texttt{GALAXY} (object touches a pixel in a large galaxy), or \texttt{CLUSTER} (object touches a pixel in a globular cluster).\footnote{See \url{https://www.legacysurvey.org/dr9/bitmasks/} for a detailed description of each maskbit.} These were chosen to be consistent with the nominal DESI QSO SV target selection \citep{QSO_desi}. We then applied a morphology cut, requiring either point sources (``PSF'' morphology) or a small relative $\chi^2$ difference between the PSF and more extended morphological models (defined in \citealt{QSO_desi}) from the DR9 legacy imaging \citep{dey,schlegel_desi}. 

To ensure we minimize non-QSO contaminants in our sample, we required a signal-to-noise (SNR) cut of $>$\,$3$ in the \textit{WISE} $W1$, $W2$, and $W3$ bands and applied the \cite{mateos} ``AGN wedge'' (a $W1-W2$ and $W2-W3$ colour-colour selection). We chose to use the \cite{mateos} wedge rather than more simple, liberal MIR cuts (e.g., \citealt{stern12}; $W1-W2\geq0.8$, Vega) in order to optimize the number of reliable QSOs observed by our SP, in addition to obtaining accurate $W3$ fluxes used to calculate the $6$\,$\upmu$m luminosity (see Section~\ref{sec:radio_data}). However, this may result in fewer sources observed which have MIR colours that lie outside of the AGN wedge (in particular, exotic QSOs with more extreme MIR colours or higher redshift QSOs for which the AGN wedge is known to be less reliable). After applying the MIR cuts, we then imposed an $r$-band magnitude cut of $r$\,$<$\,$23.5$\,mag (matching the more liberal cut used for the SV QSO selection). We further removed any object that falls within the region of $g-r$ versus $r-z$ colour-colour space that is the predominant colour selection for nominal DESI QSOs ($g-r$\,$<$\,$1.3$ \& $r-z$\,$>$\,$-0.4$ \& $r-z$\,$<$\,$1.1$; see Fig.~\ref{fig:colour_colour}), in order to expand the DESI QSO colour space. This resulted in predominantly red objects, although, as noted earlier, the selection also includes extremely blue objects (e.g., $\sim$\,12.3 per~cent of our secondary program QSOs have an $r-z<-0.4$; see Fig.~\ref{fig:colour_colour}). The nominal QSO selection includes a Random Forest selection that will also target QSOs outside this colour region, although these objects will be rare; see Fig.~\ref{fig:colour_colour2} and \citealt{yeche}. The number of objects that both satisfy these criteria, and were specifically targeted by our program, is 3038.\footnote{Note: an additional 25 objects were targeted by our program but no longer meet our photometric selection criteria due to updated photometry. We do not include these 25 in the final DESI SP sample.} It is worth noting that if this program continues for the full five years of the DESI survey, we expect over $\sim$\,8 times this number of SP targets to be observed. 

\subsection{Spectral classification}\label{sec:method_VI}

The DESI spectrograph covers an observed wavelength range of $\sim$\,3600--9800\,\AA, with a spectral resolution of $\sim$\,2000--5000 (blue to red). In order to measure the redshift and determine the spectral type (hereafter ``spectype'') of each spectrum, DESI makes use of three tools. \texttt{redrock}\footnote{\url{https://github.com/desihub/redrock}} (\citealt{redrockPaper}) is the standard DESI spectral template-fitting code that uses a set of templates to classify objects into one of three broad categories: QSO, GALAXY, and STAR. \texttt{redrock} uses a Bayesian approach and selects the best-fitting solutions based on the lowest reduced $\chi^{2}$ values. In addition to \texttt{redrock}, DESI also utilizes two ``afterburner'' codes that are applied after \texttt{redrock} is run: \texttt{QuasarNET}\footnote{\url{https://github.com/ngbusca/QuasarNET}} \citep{qn}, a neural network-based QSO classifier that updates the spectype and redshift of missed high confidence QSOs, and the Mg\,{\sc ii} afterburner, which is an emission-based code that updates the spectype of a classified GALAXY to QSO if there is significant broad Mg\,{\sc ii} emission present in the spectrum (see Section~6.2 in \citealt{QSO_desi}). We refer to the combination of these three tools as the ``modified pipeline''. 

\begin{figure}
    \centering
    \includegraphics[width=0.47\textwidth]{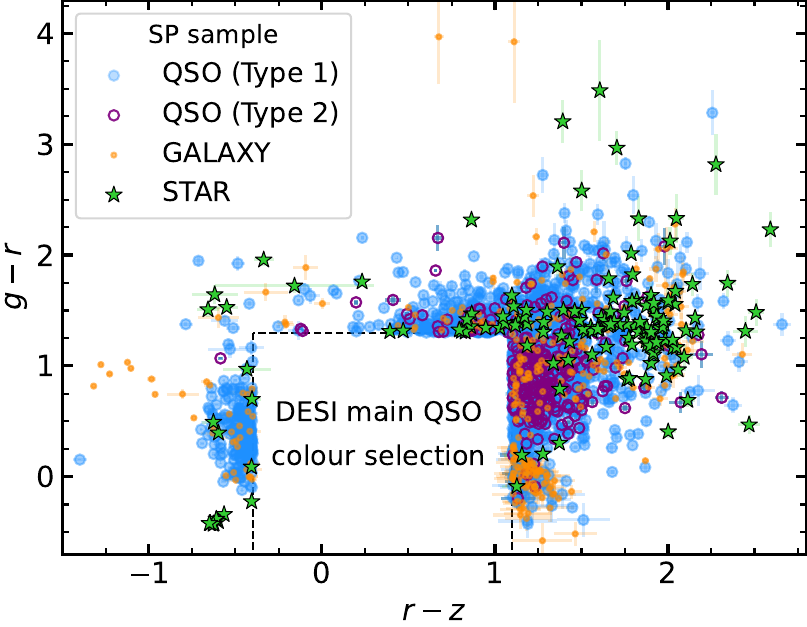}
    \caption{$g-r$ versus $r-z$ colour-colour distribution for our high-quality (VI\_quality\,$\geq$\,2.5) SP sample split by spectral type determined from VI; QSOs (Type 1: blue dots; Type 2: purple circles), galaxies (orange dots), and stars (green stars). The region of colour space that is predominantly occupied by the nominally selected DESI QSOs is highlighted; we do not target any object in this region but instead combine our sample with the nominal DESI QSOs for some of the analyses in this paper (see Section~\ref{sec:qso_sample}). It is worth noting that the nominal DESI QSO selection also utilizes a Random Forest selection that will target QSOs outside of this main colour region (see Fig.~\ref{fig:colour_colour2}).}
    \label{fig:colour_colour}
\end{figure}

On the basis of visual inspection (VI), the modified pipeline achieves a good redshift ($\Delta v$\,$<$\,$3000$\,km\,s$^{-1}$)\footnote{$\Delta v$\,$<$\,$3000$\,km\,s$^{-1}$ is equivalent to a $\Delta z$\,$=$\,$|z_{\rm VI}-z_{\rm pipeline}|/(1+z_{\rm VI})$\,$<$\,0.01, where $z_{\rm VI}$ is the redshift defined from visual inspection.} and spectral type purity of 99 per~cent and 94 per~cent, respectively for the main QSO survey (see Tables 8 and 9 in \citealt{VI1}). However, from a VI categorization of QSOs with incorrect redshifts and/or incorrect spectypes from the modified pipeline, \cite{VI1} showed that many were found to either have: 1) a very red continuum, 2) strong host-galaxy features, or 3) strong broad absorption features.\footnote{It is worth noting that improved QSO \texttt{redrock} templates are currently being developed. For more details, see \cite{brodzeller}.} Therefore, our DESI SP sample, the majority of which have very red optical colors (Fig.~\ref{fig:colour_colour}) in addition to pushing to fainter magnitudes than the main DESI QSO selection ($r$\,$<$\,$23$\,mag), is likely to result in the sort of objects that the DESI pipeline struggles to classify. Due to this, we undertook complete VI of our SP sample of 3038 objects in order to retrieve accurate redshifts and spectypes. Our VI approach is based on that described in Section~2.2 of \cite{VI1} and is summarized below. 

Each of the 3038 objects in the SP sample are evaluated using the \texttt{Prospect}\footnote{\url{https://github.com/desihub/prospect}} visualization tool; for a detailed description of the \texttt{Prospect} tool, see \cite{lan}. \texttt{Prospect} displays the un-smoothed spectrum, along with the noise spectrum and an indication of where the three arms of the DESI spectrograph overlap. Available tools on the page allow the user to smooth the spectrum and display common absorption and emission lines. The nine best-fitting \texttt{redrock} template solutions are also displayed and the best four solutions are available to overlay on the spectrum. The visual assessor is then able to manually: 1) adjust and correct the redshift, 2) correct the spectype, 3) add a specific comment, 4) flag any bad data, and 5) assess the quality of the spectrum. Examples of specific comments that were added include flagging Type 2 AGN, BALQSOs, and spectra that included two sources (i.e., those with two redshift solutions). A quality class (VI\_quality) of 0 to 4 is provided by the visual assessor for each spectrum, where 0 is the worst quality and 4 is the best quality (the specific meaning of each VI\_quality is detailed in \citealt{VI1}). 
Every spectrum is visually assessed by two visual assessors and then the files are merged by a VI lead (D. M. Alexander for this dataset), who reassess spectra with significant disagreements between the two assessors and then decides on the final classification and/or redshift.\footnote{Significant disagreements include assigning different spectypes, redshifts that differ by $>3000$\,kms$^{-1}$, and quality flags that differ by $>1$. The majority of disagreements arise for the faintest magnitude sources that are often considered low quality, and therefore are not included in our final sample.} The resulting spectrum is considered ``high quality'' with a robust redshift and spectype if the average VI quality class is $\geq2.5$ (i.e., the lowest quality spectra that are included in our high-quality sample are those where one assessor assigns a quality flag of 2 and the other assigns a 3).

\begin{figure}
    \centering
    \includegraphics[width=0.45\textwidth]{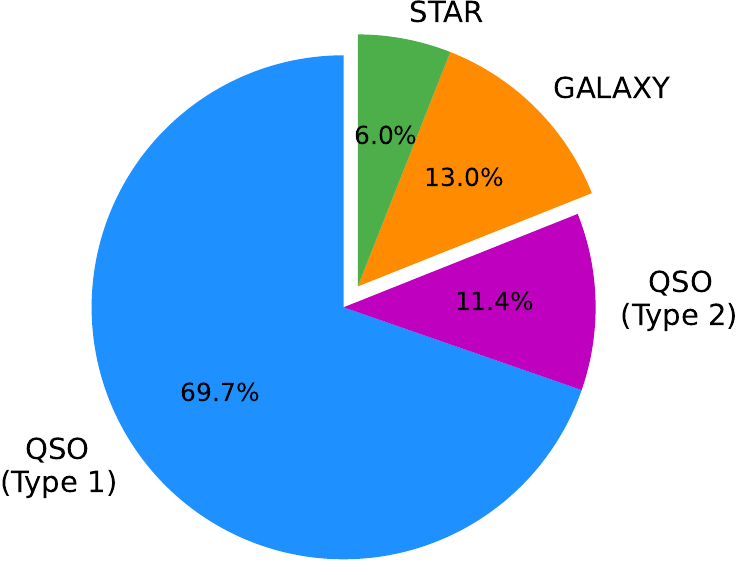}
    \caption{Pie chart showing the fraction of high-quality (VI quality\,$\geq$\,$2.5$) QSO (Type 1 and 2), STAR, and GALAXY spectral classifications from the VI of our DESI SP sample. The number of high-quality objects for each spectype in our SP sample is displayed in Table~\ref{tab:spec_tab}.}
    \label{fig:spec_pie_VI}
\end{figure}

\begin{table}
    \centering
    \begin{tabular}{cc}
        \hline 
        \hline
        Spectral type & Percentage of high-quality sources \\
        \hline
        QSO (Type 1) & 69.7 per~cent (1852/2658) \\
        QSO (Type 2) & 11.4 per~cent (302/2658) \\
        GALAXY & 13.0 per~cent (345/2658) \\
        STAR & 6.0 per~cent (159/2658) \\
         \hline
         \hline
    \end{tabular}
    \caption{Table displaying the percentage and number of high-quality (VI quality\,$\geq$\,$2.5$) objects in our DESI SP sample, in addition to removing systems with two objects in the same spectra, that were classified as QSO (Type 1 and 2), GALAXY, or STAR from the VI. Overall, we find $\sim$\,$70$ per~cent of the high-quality sources are Type 1 QSOs, which is consistent with the fraction found from the main DESI QSO selection (see Figure~4 in \citealt{VI1}).}
    \label{tab:spec_tab}
\end{table}

From visually inspecting our DESI SP sample, we found that $\sim$\,$89$ per~cent (2700/3038) of the DESI spectra have high quality (VI quality $\geq$\,$2.5$) redshifts and spectypes. Out of these high-quality objects we found that $\sim$\,1.6 per~cent (42/2700) had two redshift solutions; inspecting the images confirmed that there were indeed two objects within the same fibre, and we therefore removed these objects from our analyses. Excluding the low-quality sources and systems with two objects in the same spectrum, we found that $\sim$\,$70$ per~cent (1852/2658) of our SP sample are confirmed as Type 1 QSOs, displaying clear broad emission lines in their optical spectra; this is consistent with the expected fraction of high-quality QSOs from the main survey QSO targets ($\sim$\,$76$ per~cent; see Figure~4 in \citealt{VI1}). Additionally, $\sim$\,11 per~cent (209/1852) of the QSOs were flagged as BALQSOs, displaying broad absorption blueward of C\,{\sc iv} (e.g., \citealt{petley}). We also found that $\sim$\,$11$ per~cent (302/2658) of the high-quality spectra are Type 2 AGN, displaying strong [Ne\,{\sc v}]$\uplambda\uplambda$3346,3426 but no broad lines, $\sim$\,$13$ per~cent (345/2658) are galaxies (Fig.~\ref{fig:type2_comp}), and $\sim$\,$6$ per~cent (159/2658) are stars (Fig.~\ref{fig:star_comp}), consistent with the main DESI QSO survey \citep{VI1}; see Table~\ref{tab:spec_tab} and Fig.~\ref{fig:spec_pie_VI}. This demonstrates that the \cite{mateos} AGN wedge is efficient at selecting QSOs and that our program is indeed targeting additional QSOs beyond the main DESI survey, rather than introducing a large number of non-QSO contaminants.

Fig.~\ref{fig:colour_colour} displays the optical colour-colour distribution for the different spectral classes within our secondary sample.\footnote{Note: 66 systems were removed due to imaging issues (e.g., image artifacts, multiple objects in the same fibre, bright nearby stars, edge effects) that resulted in poor photometry in one of the bands, and therefore unrealistic colours.} We find that $\sim$\,87, $\sim$\,84, and $\sim$\,81 per~cent of the QSOs, galaxies, and stars have red optical colours ($g-r$\,$>$\,$1.3$\,mag or $r-z$\,$>$\,$1.1$\,mag), respectively. The number of sources for each spectype identified from our VI is displayed in Table~\ref{tab:spec_tab}. For a detailed comparison of the performance of the modified pipeline compared to the VI results, see Appendix~\ref{sec:appendix_MP}. In this paper we focus on the QSOs in our sample (Fig.~\ref{fig:flow2}); for more discussion on the galaxies and stars within our DESI SP sample, see Appendix~\ref{sec:properties_stack}.

\section{Data and methods}\label{sec:final_qso}
In this section, we compare the properties of the QSOs selected by our secondary program to those selected by the nominal DESI QSO programs (Section~\ref{sec:qso_sample}), our spectral stacking method (Section~\ref{sec:comp}), our dust fitting method (Section~\ref{sec:dust_desi}), the radio data utilized in this paper (Section~\ref{sec:radio_data}), and the construction of the combined QSO sample used for the main analyses in this paper (Section~\ref{sec:radio_desi}).

\subsection{How does our SP extend the nominal DESI QSO selection?}\label{sec:qso_sample}

\begin{figure}
    \centering
    \includegraphics[width=0.48\textwidth]{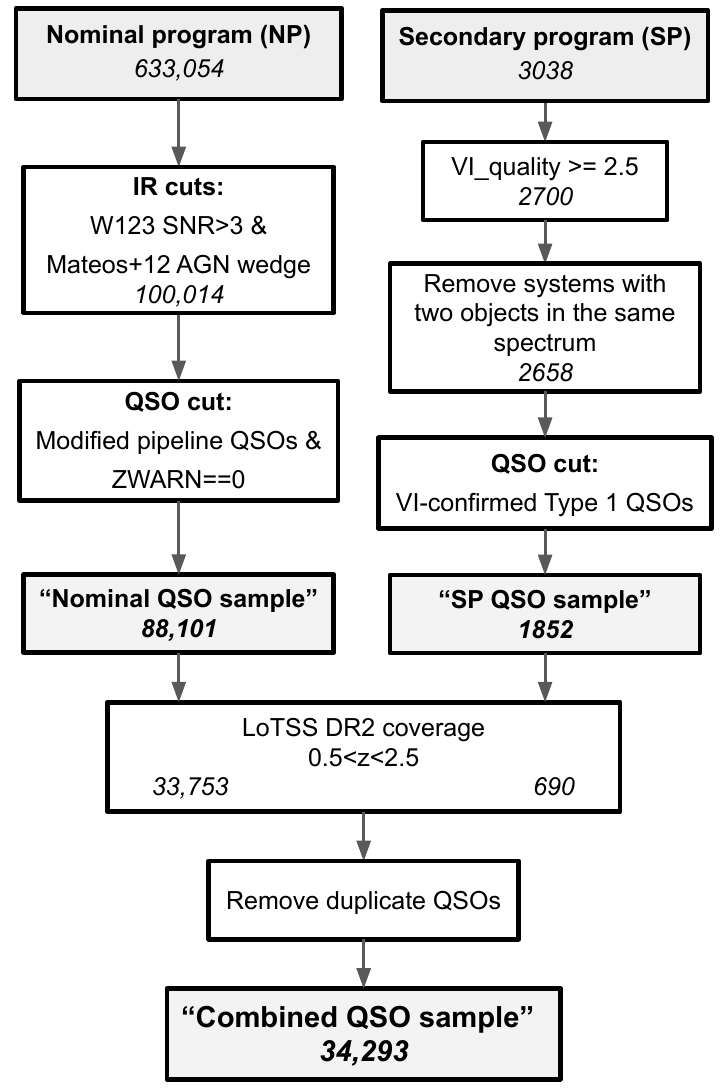}
    \caption{Flow chart illustrating the selection process for the combined QSO sample. We restricted the SP sample (see Fig.~\ref{fig:flow} for sample selection) to high-quality (VI\_quality\,$\geq$\,$2.5$; see Section~\ref{sec:method_VI}) objects and removed systems with two objects in the same spectrum (i.e., two redshift solutions). We then selected the visually confirmed QSOs, resulting in 1852 ``SP QSOs''. We selected objects that were targeted by the nominal QSO program and then applied the same IR cuts used in our SP sample selection. We then selected QSOs based on the spectype from the modified pipeline and also applied a ZWARN cut to ensure the redshifts were robust. The resulting ``nominal QSO sample'' consists of 88\,101 QSOs. Restricting both samples to the LoTSS DR2 coverage and a redshift range of $0.5$\,$<$\,$z$\,$<$\,$2.5$, we then combined both QSO sub-samples, resulting in 34\,293 QSOs that form the ``combined QSO sample''. An electronic table containing the DESI \texttt{TARGETID} for the full combined QSO sample, and additionally the RA, Dec, redshift, spectype, $L_{\rm 6\,\upmu m}$, and $L_{\rm 144\,MHz}$ for the combined QSO sample included in the DESI Early Data Release \citep{sv}, is available online.}
    \label{fig:flow2}
\end{figure}

For the analyses in this paper we focus on the broad line QSOs in our SP sample (excluding the 306 Type 2 AGN; see Table~\ref{tab:spec_tab}). We also require a comparison QSO sample in order to assess how our SP sample extends the nominal DESI QSO population. We refer to the QSOs from the nominal DESI QSO population as ``nominal QSOs'' and the QSOs from our SP as ``SP QSOs''. Our selection process for the nominal and SP QSO samples is displayed in Fig.~\ref{fig:flow2}, which we describe below.

\begin{figure}
    \centering
    \includegraphics[width=0.45\textwidth]{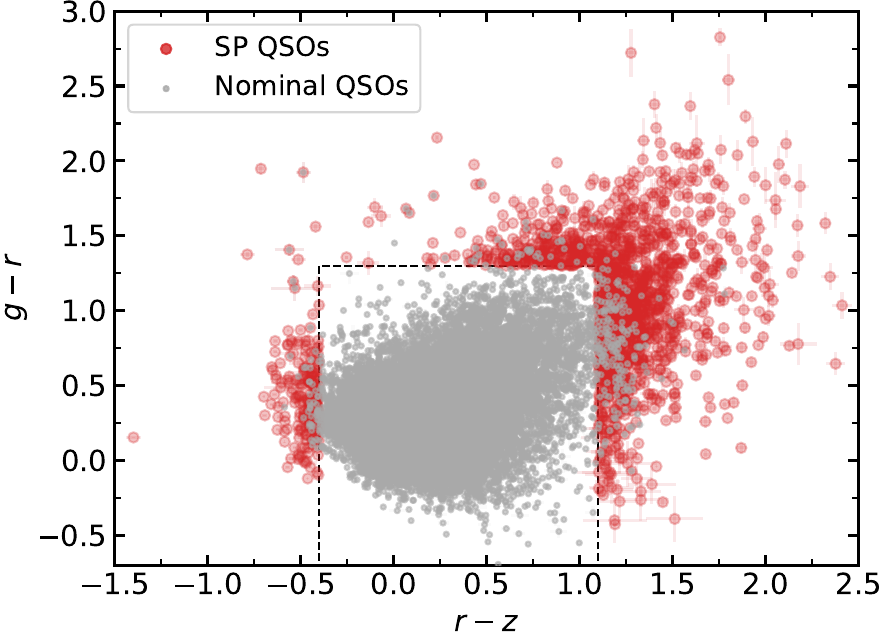}
    \includegraphics[width=0.45\textwidth]{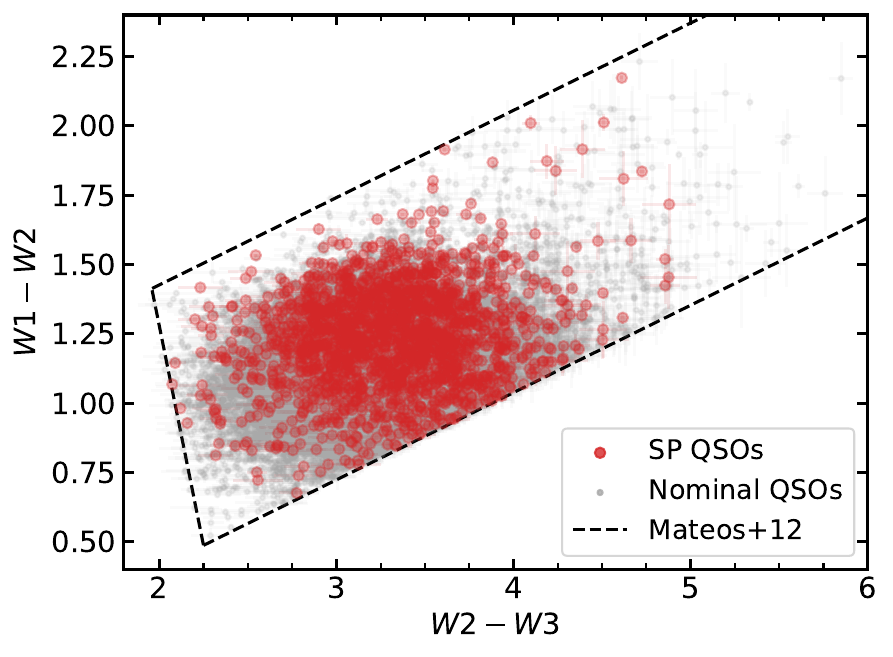}
    \caption{(Top) similar to Fig.~\ref{fig:colour_colour}, but now displaying the SP (red) and nominal (grey) QSO samples (see Fig.~\ref{fig:flow2}). Only $\sim$\,3 per~cent (2367/88\,101) of the nominal QSO sample lie outside the colour region indicated by the dashed lines; these QSOs were likely selected by the Random Forest QSO selection. Our SP QSO sample provides 1107 additional unique (i.e., solely targeted by our program) QSOs to the nominal DESI QSO survey, the majority of which display very red colours (i.e., large $g-r$ and/or $r-z$ colours). (Bottom) \textit{WISE} $W1-W2$ versus $W2-W3$ for the SP (red) and nominal (grey) QSO samples.}
    \label{fig:colour_colour2}
\end{figure}

To construct the nominal QSO sample, we first selected objects that were targeted by the nominal DESI QSO program (bitmask\,$=$\,\texttt{QSO}). We then applied the same MIR cuts that were used to form our SP sample (Section~\ref{sec:sample_desi}; SNR\,$>3$ in the $W123$ bands and satisfying the \cite{mateos} AGN wedge; see Fig.~\ref{fig:colour_colour2}). Finally, we restricted the selection to objects that were classified as QSOs by the modified pipeline and apply ZWARN\,$==0$ to ensure the redshifts are robust. The resulting nominal QSO sample consists of 88\,101 QSOs. To construct the SP QSO sample, we selected the visually confirmed QSOs from our SP sample (Section~\ref{sec:method_VI}) which had a VI\_quality\,$\geq2.5$ (i.e., had a robust spectype and redshift) and additionally removed the Type 2 AGN (no broad lines present) and systems with two objects in the same spectrum; this resulted in 1852 QSOs in our SP QSO sample

Due to re-observations, many DESI targets will obtain multiple spectra. In these cases, for our analyses we use spectra that coadd all observations of a given target. In addition to this, since the DESI SV observations each have distinct target selections (for more details see \citealt{myers}) there may be multiple coadded spectra across the different campaigns. In these cases, we use the coadded spectrum with the highest squared template signal-to-noise ratio (TSNR2) value. The calculation of the TSNR2 value is target-class dependent and so in this work we use the TSNR2\_QSO statistic. For more details on the TSNR2 metric, see \cite{DataPaper}.

Fig.~\ref{fig:colour_colour2} displays the optical colour-colour distribution for both the nominal and SP QSO samples. We find that, as expected, the majority of the nominal QSO sample lie inside the main colour space used to select DESI QSOs, with only $\sim$\,3 per~cent (2367/88\,101) of QSOs that lie outside; these QSOs were likely selected by the Random Forest QSO selection. Our SP QSO sample expands the colour parameter space of DESI QSOs, with the majority of additional QSOs displaying very red optical colours. It is worth noting that $\sim$\,40 per~cent (745/1852) of the SP QSOs were also targeted by the nominal QSO program; these 745 QSOs met both our DESI SP selection criteria and one of the nominal QSO selections. Our DESI SP therefore provides $\sim$\,32 per~cent more QSOs that lie outside the main region of colour space, i.e., 1107 SP QSOs solely targeted by our program compared to the 2367 nominal QSOs that fall outside the main colour region (see Fig.~\ref{fig:colour_colour2}), despite being only $\sim$\,2 per~cent the size of the nominal QSO sample.

To understand how the 1852 DESI SP QSOs compare to the SDSS QSOs utilized in our previous studies, we first restricted the SP QSO sample to the SDSS footprint, reducing the sample to 1312 QSOs. We then matched to the DR16 QSO catalogue \citep{dr16} using a $1''$ search radius; this resulted in 107 matches. Therefore, only 8 per~cent (107/1312) of the SP QSO sample within the SDSS footprint have also been observed by SDSS and met the criteria used by \cite{dr16} to be included in the SDSS DR16 QSO catalog. Out of the 107 sources, we found seven with a redshift disagreement between our VI and the SDSS DR16 QSO catalogue; additional VI of these objects clarified that the SDSS redshifts were significantly incorrect. Matching the DR16 QSO catalogue with the other high-quality sources from our SP sample, we found three matches; one was classified as a galaxy and two were classified as Type 2 AGN by our VI. We further investigate these disagreements in Appendix~\ref{sec:sdss_desi}.

\begin{figure}
    \centering
    \includegraphics[width=0.47\textwidth]{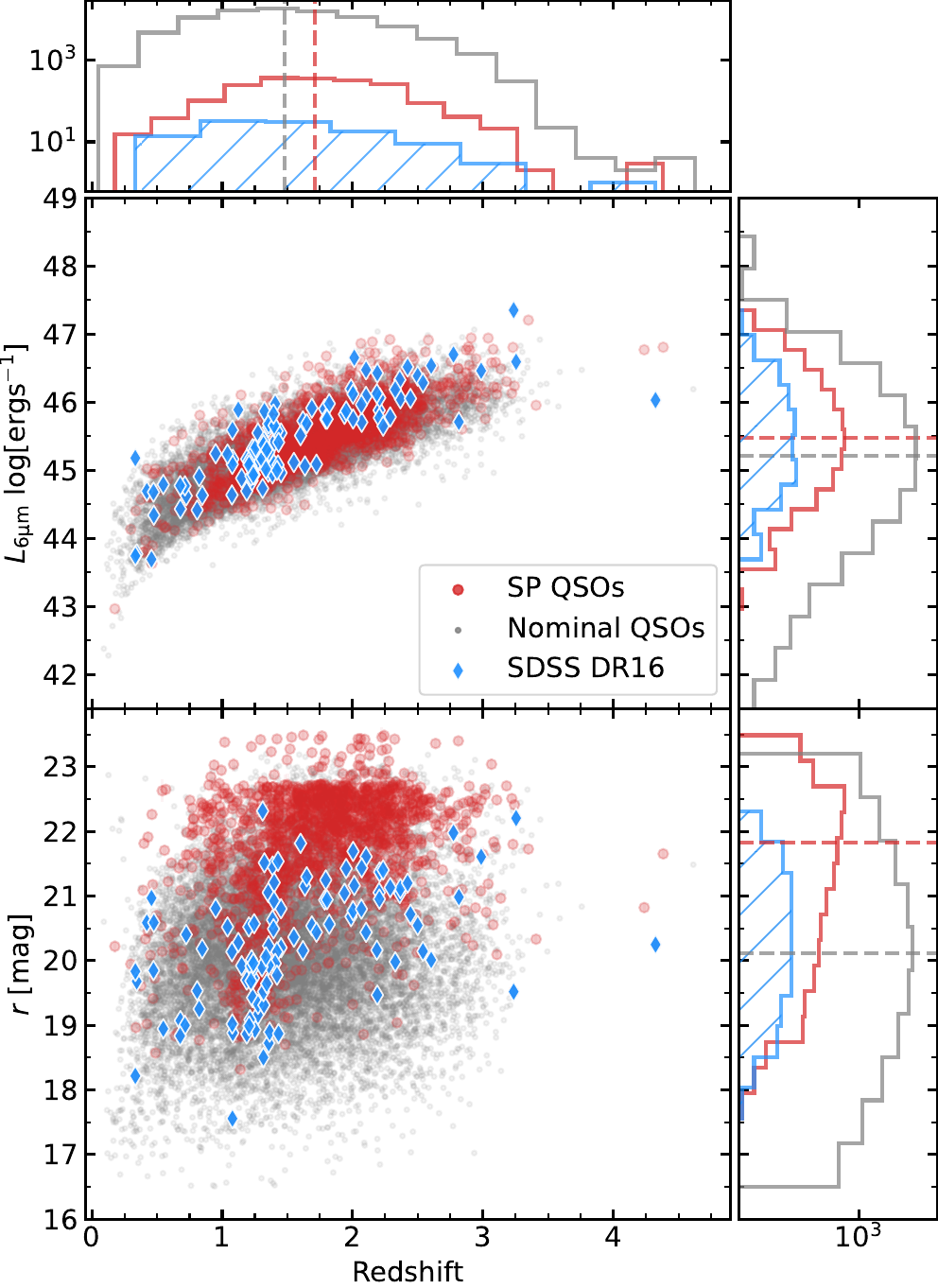}
    \caption{$L_{\rm 6\,\upmu m}$ (top) and $r$-band magnitude (bottom) versus redshift for the nominal QSO (grey dots) and SP QSO (red circles) samples. The median values for the DESI nominal and SP QSO samples are shown by the dashed lines in the adjacent histograms. The 107 SP QSOs that are also identified in the SDSS DR16 QSO catalogue are highlighted by the blue diamonds.}
    \label{fig:L6_red}
\end{figure}

Fig.~\ref{fig:L6_red} displays the $L_{\rm 6\,\upmu m}$ and $r$-band magnitude versus redshift distributions for the nominal and SP QSO samples. To compute the \textit{L}\textsubscript{6\,$\upmu$m} we used a log-linear interpolation or extrapolation of the fluxes in the \textit{W2} and \textit{W3} bands, following the approach from \cite{klindt}. 
The larger density of SP QSOs at $r$\,$\lesssim$\,$22.5$\,mag is a result of the different target prioritization during SV3 (see Section~\ref{sec:sample_desi}); SP sources {\em that were also primary targets} i.e., QSOs \citep{QSO_desi}, LRGs \citep{lrg}, or Emission Line Galaxies (ELGs; \citealt{elg}), will have had a far higher probability of being observed than SP targets that were {\em not} also primary targets. The selection of these primary targets impose brighter magnitude limits than that of our SP, and therefore the targets observed during SV3 are more likely to reside at these brighter magnitudes. The SP QSOs that were also observed by SDSS are shown by the blue diamonds. The highest redshift QSO in the SP and nominal samples are at $z$\,$=$\,$4.4$ and 4.6, respectively. On average, the SP QSOs are $\sim$\,$1.7$ magnitudes optically fainter than the QSOs in the nominal sample, demonstrating that our program is pushing to QSOs with fainter magnitudes. Applying the two-sided Kolmogorov-Smirnov (K-S) test to the redshift and $L_{\rm 6\,\upmu m}$ distributions of the nominal and SP QSO samples, we cannot rule out that the two samples are drawn from the same parent distribution at a $<1$ per~cent significance level. This also demonstrates that QSOs with fainter optical magnitudes are not necessarily intrinsically fainter; their optical emission may be relatively fainter due to dust extinction (see Section~\ref{sec:dust_desi_result}). We utilize $L_{\rm 6\,\upmu m}$ as a tracer for the intrinsic QSO luminosity in Section~\ref{sec:radio_desi_result}.

\subsection{Composite construction}\label{sec:comp}
To characterize the basic spectral properties of our QSO samples we constructed composite spectra following the approach outlined in \cite{fawcett22}; hereafter, \citetalias{fawcett22}. First, we trim the ends of the spectrum from each spectral arm to remove noisy data. We then correct for Galactic extinction, using the \cite{schlegel} map and the \cite{fitz} Milky Way extinction law, and shift to rest-frame wavelengths using the VI redshifts for the SP QSO sample and the redshifts obtained from the modified pipeline for the nominal QSO sample. Each spectrum is then re-binned to a common wavelength grid with 1\,\AA~per bin and normalized to 4500\,\AA. We then take the median composite, applying a minimum threshold of 30 spectra per bin.

\subsection{Dust extinction fitting method}\label{sec:dust_desi}
To quantify the amount of extinction present in each QSO, we fitted a blue QSO spectral template with varying amounts of dust extinction to the QSO spectra following the same method as \citetalias{fawcett22}. We briefly outline the approach here. 

For our blue QSO template, we used the VLT/\textit{X-shooter} control QSO composite from \citetalias{fawcett22}. This composite was constructed by taking the geometric mean of 28 blue QSOs at $1.45$\,$<$\,$z$\,$<$\,$1.65$ (spanning rest-frame wavelengths $1250$\,$<$\,$\uplambda$\,$<$\,$9000$\,\AA) observed with \textit{X-shooter} (Program ID 0101.B-0739; PI: Klindt). We chose to use this composite rather than constructing a blue composite from DESI spectra due to the broad wavelength coverage of \textit{X-shooter}; to achieve a similar wavelength range with DESI, we would need to use objects at different redshifts which could result in a non-physical continuum shape (see Appendix~D in \citetalias{fawcett22} for more discussion). We then masked the emission-line regions and smoothed the spectrum of the blue QSO \textit{X-shooter} composite with a Gaussian filter ($\upsigma=3$). We fitted this unreddened template to the QSO spectra using a least-squares minimization code, which varied the normalization (avoiding emission-line regions) and $E(B-V)$ parameter (ranging from $-1.0$\,$<$\,$E(B-V)$\,$<$\,$2.5$\,mag\footnote{We note that, even given the natural spread in the power-law slopes for unobscured QSOs, a QSO with an $E(B-V)$\,$=$\,$-1$\,mag is physically impossible. We chose this lower limit to reduce boundary effects in the dust extinction fitting code. For the analyses in this paper, we apply at $E(B-V)$\,$>$\,$-0.1$\,mag cut to esnure we only use reliable fits.}), using a simple power-law (PL; $A_V\propto\uplambda^{-1}$, $R_V=4.0$; \citealt{fawcett22}) dust extinction curve which was found to describe the dust in dust-reddened QSOs more effectively than the commonly used Small Magellanic Cloud (SMC; \citealt{SMC}) extinction law \citep{fawcett22}.\footnote{The \textit{X-shooter} composites and a movie describing our basic fitting approach can be found online: \url{https://github.com/VFawcett/XshooterComposites}} The measured $E(B-V)$ was allowed to be negative to account for the fact that the \textit{X-shooter} blue QSO composite used for the fitting will contain a small amount of dust and, therefore, some DESI QSOs will be bluer than the composite. However, QSOs with negative $E(B-V)$ values should be interpreted with caution; they represent the natural spread in the power-law slope for unobscured QSOs (e.g., \citealt{richards}), but do not have any physical interpretation.

In order to determine the quality of the fits, we calculated the Mean Absolute Deviation (MAD\,$=$\,$\frac{1}{n}\sum|\rm X-\upmu|$; where $n$ is the sample size, X is difference between the model and data, and $\upmu$ is the mean of the difference). For this work, we choose a conservative cut of MAD\,$<$\,$4\times10^{-5}$ to define whether a fit is good or not. Although this will cut down our sample substantially, it should ensure the results in this paper are robust (see Appendix~\ref{sec:test} for a comparison of how different MAD cuts affect our results). 

It has previously been noted that QSOs with low luminosities (log\,$L_{\rm 6\upmu m}\lesssim45$\,erg\,s$^{-1}$) and redshifts ($z$\,$\lesssim$\,$1$) can have a substantial contribution of light from the host galaxy \citep{shen11,calistro,fawcett_thesis}. Therefore, for the QSOs at $z\leq1$, we first fitted and subtracted a host-galaxy model from the spectrum to avoid potential stellar emission biasing dust extinction estimates towards the QSO. To fit the host-galaxy component, we used the publicly available code \texttt{PyQSOFit}\footnote{\url{https://github.com/legolason/PyQSOFit}}, following the approach taken in \citetalias{fawcett22}. For each source, we first smoothed the spectra with a 10 pixel box-car and then globally fitted both the continuum and emission lines of the entire spectrum. To fit the continuum, we used a power-law, Balmer continuum (BC), third-order polynomial, \ion{Fe}{ii} continuum, and host-galaxy components. Following the same approach as \citetalias{fawcett22}, we chose to apply the \cite{Verner_2009} \ion{Fe}{ii} template for $\uplambda$\,$>$\,$2000$\,\AA~(obtained via private communication), combined with the \cite{vest} template for $\uplambda$\,$<$\,$2000$\,\AA~(available on \url{https://github.com/legolason/PyQSOFit/blob/master/pyqsofit/fe_uv.txt}), to give a continuous template from 1000--12000\,\AA. The host galaxy and QSO contributions are decomposed using principal component analysis (PCA; \citealt{yip,yipb}). The PCA method is based on the assumption that the observed QSO spectrum is a combination of two independent sets of eigenspectra taken from pure galaxy and pure QSO samples. We used 5 galaxy and 20 QSO PCA components (following a similar approach as \citealt{rakshit}), and adopted the stellar host model from \cite{bruzual}. We used this model instead of the default \cite{yip} model in \texttt{PyQSOFit} due to the wider wavelength coverage. A host-galaxy component is only included if $>$\,$100$ pixels in the host-galaxy template are non-negative; we note the lower resolution of the \cite{bruzual} compared to the \cite{yip} template at higher redshifts may lead to fewer spectra successfully fitted by a host-galaxy template. Overall, we found that $\sim$\,$10$ per~cent (97/948) of the SP QSO sample at $z\leq1$ could successfully be fit with a host-galaxy component. Each of the other continuum components is described in detail in \citetalias{fawcett22}. For examples of our dust fitting method, see Fig.~\ref{fig:dust_fit_example}. 

Due to the limited wavelength range of our chosen blue composite ($1250$\,$<$\,$\uplambda$\,$<$\,$9000$\,\AA), our dust extinction fitting method is less effective for sources at $z$\,$>$\,$2.5$ since this corresponds to a large portion of the DESI spectrum with rest-frame wavelengths $<$\,$1250$\,\AA. The fitting procedure is most effective at $z$\,$<$\,$2.0$, where the whole DESI wavelength range is covered by the composite. However, to maximize the source statistics for our dust extinction analyses, we include QSOs at $2.0$\,$<$\,$z$\,$<$\,$2.5$ that have good fits.

\subsection{Radio data}\label{sec:radio_data}

In this work we utilize radio data from the LOFAR Two Metre Sky Survey (LoTSS) Data Release 2 (LDR2; \citealt{shim})\footnote{\url{https://lofar-surveys.org/surveys.html}}. This survey covers 5740 square degrees at 144\,MHz, with a median sensitivity of 83\,$\upmu$Jy/beam and $6''$ resolution. LDR2 contains 4\,496\,228 sources with a $>$\,$5$\,$\sigma$ detection. At equivalent frequencies, for sources with typical synchrotron spectra, LDR2 is $\sim$\,$9$ times deeper than the Faint Images of the Radio Sky at Twenty-centimeters (FIRST; \citealt{becker}) 1.4\,GHz survey; this makes LDR2 ideal to analyze the radio properties of DESI QSOs, which can have modest AGN luminosities.

To determine whether a QSO is `radio-quiet' or `radio-loud', we adopted the same ``radio-loudness'' parameter ($\mathcal{R}$) as that first used in \cite{klindt}, and adapted for 144\,MHz data in \cite{rosario}, defined as the dimensionless~quantity:
\begin{equation}\label{eq:radio}
\mathcal{R}=\textrm{log\textsubscript{10}}\frac{\textit{L}\textsubscript{144\,MHz}}{\textit{L}\textsubscript{6\,$\upmu$m}} .\
\end{equation}
The $L_{\rm 144\,MHz}$ is calculated using the methodology described in \cite{alex2003}, assuming a uniform radio spectral index of $\alpha$\,$=$\,$0.7$ for the $K$-correction. We used the rest-frame 6\,$\upmu$m luminosity to define radio loudness, rather than optical luminosity which is used in most classical definitions (e.g., \citealt{kellermann}), since \textit{L}\textsubscript{6\,$\upmu$m} is less sensitive to extinction. For example, even at the most extreme $E(B-V)$ for an optically selected QSO ($E(B-V)=1$\,mag) at $z=1.5$, the difference in \textit{L}\textsubscript{6\,$\upmu$m} due to loss of flux by dust extinction at rest-frame 6\,$\upmu$m is $\sim$\,0.15 dex. We scaled the radio-loud/radio-quiet threshold from 1.4\,GHz ($\mathcal{R}$\,$=$\,$-4.2$) to our 144\,MHz data by assuming a canonical spectral index of $\alpha$\,$=$\,0.7 ($S_{\nu}\propto \nu^{-\alpha}$).\footnote{\label{note:spectral}It should be noted that there will be large uncertainties on the radio spectral slope; for example, flatter spectral indices have been found for fainter QSOs (e.g., \citealt{gloudemans}). The value of $\alpha$\,$=$\,$0.7$ was chosen based on the average 144\,MHz--1.4\,GHz radio spectral index found for SDSS QSOs in \cite{rosario}.} This corresponded to a boundary at $\mathcal{R}$\,$=$\,$-4.5$, which is broadly consistent with the classical radio-quiet/radio-loud threshold (often defined using a 5\,GHz-to-2500\,\AA~flux ratio; \citealt{kellermann}), but is less susceptible to obscuration from dust (see \citealt{klindt} for full details).

\subsection{Constructing the combined QSO sample}\label{sec:radio_desi}

For the scientific analyses in this paper, we constructed a combined QSO sample, which included both our SP and nominal QSO samples in order to maximize the source statistics. We first restricted our nominal and SP QSO samples to the LDR2 survey area using the Multi-Order Coverage map (MOC) provided on the LOFAR website.\footnote{\url{https://lofar-surveys.org/dr2_release.html}} We then restricted the sample to redshifts $0.5$\,$<$\,$z$\,$<2.5$; the lower bound is to ensure minimal contamination from the host galaxy and the upper bound is due to a limitation with our dust fitting method (see Section~\ref{sec:dust_desi}). This resulted in 690 and 33\,753 QSOs in the SP and nominal samples, respectively. We then matched to the LDR2, adopting a $5''$ search radius which resulted in $\sim$\,$19$ per~cent (6389/33\,753) and $\sim$\,$31$ per~cent (216/690) radio-detected QSOs in the nominal and SP samples, respectively (we further explore the radio detection fractions in Section~\ref{sec:radio_desi_result}). Finally, we combined both the restricted nominal and SP QSO samples (removing duplicate QSOs that belong to both samples), resulting in our ``combined QSO sample'' which consists of 34\,293 QSOs (see Fig.~\ref{fig:flow2}). 

To estimate the fraction of false matches and objects missed due to radio emission extended beyond $5''$, we utilized the LoTSS DR1 radio catalogue (LDR1; \citealt{lotss}), with optical counterparts from \cite{lofar_zoo}.\footnote{At the time of writing, the optical and IR identification catalogue for LDR2 has not yet been publicly released.} The catalogue from \cite{lofar_zoo} was created using a combination of likelihood-ratio matching and visual identification, resulting in optical and/or IR counterparts for $\sim$\,73 per~cent of the LDR1 radio sources. For our false matching estimate, we will use the optical/IR positions of the LDR1 sources as the ``true'' positions. Restricting the nominal QSO sample to the much smaller region of LDR1 reduced the sample to only 3057 sources. Adopting the same $5''$ search radius as for LDR2, we matched this reduced sample to the radio positions in LDR1, resulting in 645 matches. We then repeated the matching, but instead adopting a 0$\farcs$75 search radius (following \citealt{rosario}) and matching to the positions of the optical counterparts from the value-added catalogue, resulting in 648 matches which represent our ``truth'' sample. From this, we were able to determine an estimate for the fraction of false matches by exploring how many of the 645 radio matches within $5''$ were not also included in our truth sample; from this we found a false match fraction of $\sim$\,1 per~cent (8/645). Similarly, we were able to determine the fraction of missed matches by exploring how many of the truth sample were not included in the sample of 645; from this we found that $\sim$\,2 per~cent (11/645) of sources are missed, most likely due to radio emission extended beyond $5''$. Although the false match fraction calculated here is greater than that found for FIRST (using a $10''$ matching radius: $\sim$\,0.2 per~cent; \citealt{klindt}), which is at a comparable resolution to LoTSS (FIRST: $5''$; LoTSS: $6''$), the sensitivity of LoTSS is $\sim$\,9 times deeper than FIRST at equivalent frequencies, assuming a spectral index of $\alpha$\,$=$\,$0.7$. We would therefore expect a higher false match fraction from LoTSS due to the higher density of faint sources within a given region. Additionally, our current estimations are limited by source statistics. Our future DESI papers will benefit from the upcoming optical and IR cross-matched LDR2 catalogue for more accurate cross-matching and also a larger DESI SP QSO sample.

\section{Results}\label{sec:results}
Our results are based on the combined QSO sample (see Fig.~\ref{fig:flow2}) that includes QSOs selected from our secondary target program (Section~\ref{sec:sample_desi}) and QSOs that were targeted by the nominal DESI QSO program, restricted to the LoTSS DR2 coverage area (to assess radio detection fractions). In the following sections, we explore the line-of-sight dust extinction towards the QSOs (Section~\ref{sec:dust_desi_result}) and the radio properties of the QSOs as a function of dust extinction (Section~\ref{sec:radio_desi_result}).

\subsection{Quantifying the line-of-sight dust extinction}\label{sec:dust_desi_result}

Under the assumption that dust is the main cause of reddening in reddened QSOs, we quantified the amount of dust extinction along the line-of-sight by fitting the spectra of the combined QSO sample with a dust-reddened VLT/\textit{X-shooter} blue QSO composite (see Section~\ref{sec:dust_desi}).

\begin{figure*}
    \centering
    \includegraphics[width=0.93\textwidth]{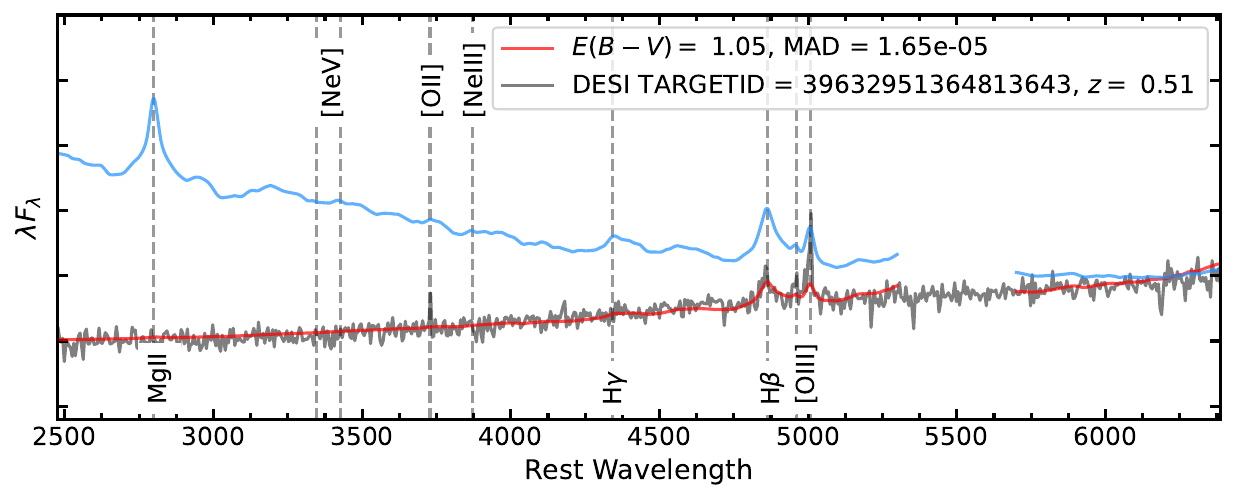}
    \caption{An example of a good dust extinction fit, based on the method described in Section~\ref{sec:dust_desi}, for a high $E(B-V)$ QSO. The blue QSO composite (from \citetalias{fawcett22}) before and after applying extinction is displayed by the blue and red lines, respectively. The DESI QSO spectrum is displayed by the grey line. The gap in the blue composite between $\uplambda$\,$\sim$\,5300--5700\,\AA~is due to the removal of strong telluric features (see \citetalias{fawcett22} for more details). Additional examples of dust extinction fits, including QSOs with poor fits, can be found in the online supplementary material.}
    \label{fig:dust_fit_example}
\end{figure*}

Overall, 58 per~cent (20\,031/34\,293) of the combined QSO sample spectra are well fitted (MAD\,$<$\,$4\times10^{-5}$; see Section~\ref{sec:dust_desi}) by the dust-reddened composite. Fig.~\ref{fig:dust_fit_example} displays the best fitting solution for one of the most extreme cases, with a measured $E(B-V)$\,$=$\,$1.05$\,mag ($A_V$\,$\sim$\,$4$\,mag); for this source the Mg\,{\sc ii} emission line is completely obscured, although broad H\,$\beta$ is still present indicating that it is a QSO. Inspecting the QSOs that are not well fit with a dust extinction curve, we found that they tend to have bluer optical colours (median $g-r=0.26$ and 0.21\,mag for the QSOs with good and bad fits, respectively), fainter optical magnitudes (median $r=20.0$ and 20.6\,mag, for the QSOs with good and bad fits, respectively), display strong absorption features, and/or reside at higher redshifts (median redshift\,$=1.22$ and $1.83$ for the QSOs with good and bad fits, respectively); as expected based on the lower effectiveness of our fitting procedure at $z$\,$>$\,$2$ due to the limited wavelength range of our chosen blue composite (see Section~\ref{sec:dust_desi}). We therefore expect the QSOs with good fits to be slightly biased towards either the more luminous QSOs and/or those with higher dust extinctions. Additionally, from visually assessing the bluest QSOs ($E(B-V)$\,$<$\,$-0.1$\,mag) with a MAD\,$<$\,$4\times10^{-5}$, we found the fits to be unreliable (i.e., the extinction of the QSO compared to the blue composite appears to be consistent with zero, but the measured $E(B-V)$ is clearly underestimated), again highlighting the limitation of our fitting towards blue QSOs. However, due to the significant sample size of the bluest QSOs we expect this to have little effect on our results (see Appendix~\ref{sec:test}). For the following analyses, we restrict to QSOs with good dust extinction fits and also with a measured $E(B-V)$\,$>$\,$-0.1$\,mag. Examples of QSOs with good and bad dust extinction fits can be found in the online supplementary material.

\begin{figure}
    \centering
    \includegraphics[width=0.45\textwidth]{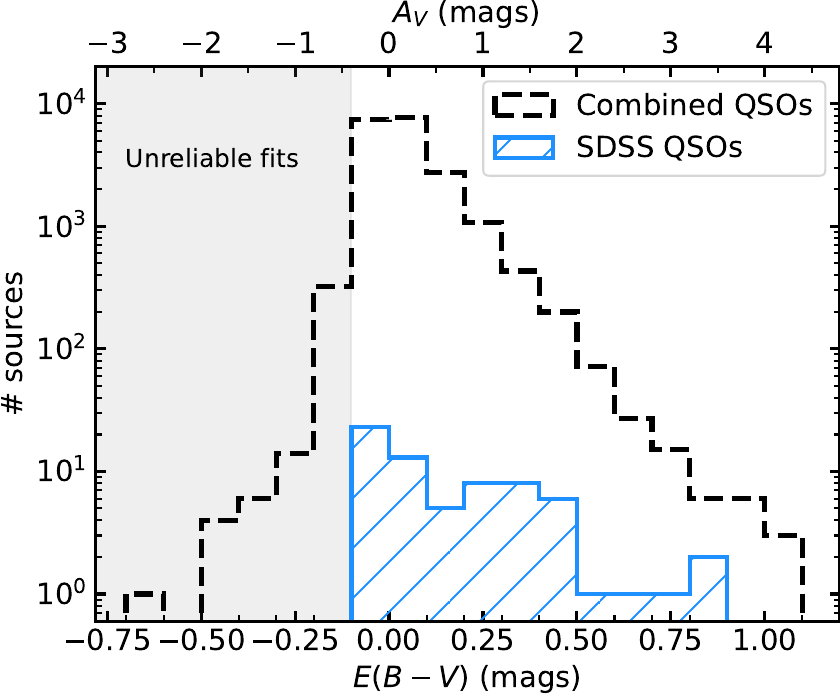}
    \caption{$E(B-V)$ distribution for the combined QSO sample (black dashed line) and SP QSOs that are were also observed by SDSS (hatched blue region), restricted to the sources with good $E(B-V)$ fits. With DESI, we can now observe QSOs with dust extinction values up to $E(B-V)$\,$\sim$\,1 ($A_V$\,$\sim$\,4).} 
    \label{fig:av_hist_all}
\end{figure}

\begin{figure}
    \centering
    \includegraphics[width=0.5\textwidth]{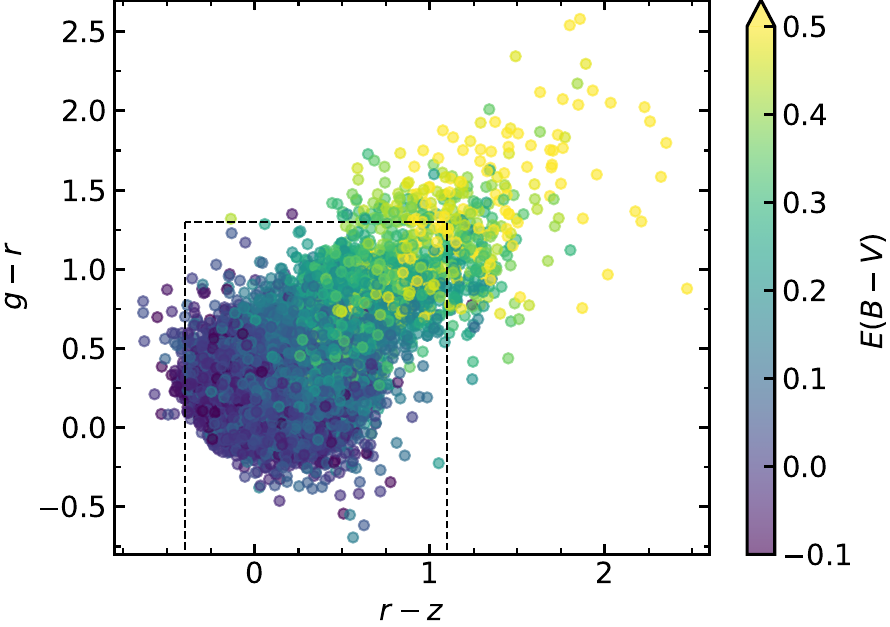}
    \caption{$g-r$ versus $r-z$ for the combined QSO sample, restricted to the QSOs with good $E(B-V)$ fits. The colourbar displays the measured $E(B-V)$ which scales with both $g-r$ and $r-z$ colour. The dashed lines represent the optical colour selection applied in our SP. This demonstrates that optical colour, and our SP, is an effective way of selecting dust-reddened QSOs.}
    \label{fig:gr_rz_colour}
\end{figure}

Fig.~\ref{fig:av_hist_all} displays the distribution of dust-extinction values for the combined QSO sample. 
The 107 SP QSOs that were additionally identified in the SDSS are also highlighted. This demonstrates that DESI will not only accumulate a more statistically significant sample of reddened QSOs but also extends to QSOs with dust extinctions beyond those observed by the SDSS. 
Fig.~\ref{fig:gr_rz_colour} displays the $g-r$ versus $r-z$ optical colour-colour space, with the colourbar representing the measured $E(B-V)$. We find that both the $g-r$ and $r-z$ colour scale with $E(B-V)$, demonstrating that optical colour is a good proxy for dust reddening. In order to maximize the highest $E(B-V)$ end of the QSO population, future QSO selections can utilize this correlation between optical colour and dust extinction in order to select the reddest QSOs.

Overall, DESI can identify QSOs with dust extinctions up to $E(B-V)$\,$\sim$\,1.0\,mag ($A_V$\,$\sim$\,$4$\,mag; see Fig.~\ref{fig:dust_fit_example} for an example fit), comparable to the much smaller sample of NIR-selected heavily reddened QSOs from \cite{ban}. This demonstrates the capability of DESI to help bridge the gap between mildly obscured SDSS red QSOs (e.g., \citealt{richards,klindt,fawcett22}), IR-selected red QSOs (e.g., \citealt{ban,glik15,glikman22}), and heavily obscured QSOs (e.g., \citealt{andonie}). See Section~\ref{sec:red_dis} for a more detailed comparison with previous red QSO studies. The measured $E(B-V)$ values for the combined QSOs included in the DESI EDR can be found in the electronic table online.

\subsection{Exploring the radio properties of dusty QSOs}\label{sec:radio_desi_result}

In our previous work, we identified fundamental differences in the radio properties of red QSOs compared to typical blue QSOs based on the shallower SDSS QSO samples \citep{klindt,rosario,fawcett20,fawcett21}. In this paper we utilize the combined QSO sample, our spectral fitting results, and LDR2 (see Sections~\ref{sec:radio_data} and \ref{sec:radio_desi}) in order to explore the radio detection fraction as a function of reddening and extend our previous radio results to more heavily reddened QSOs. 

\begin{figure*}
    \centering
    \includegraphics[width=0.75\textwidth]{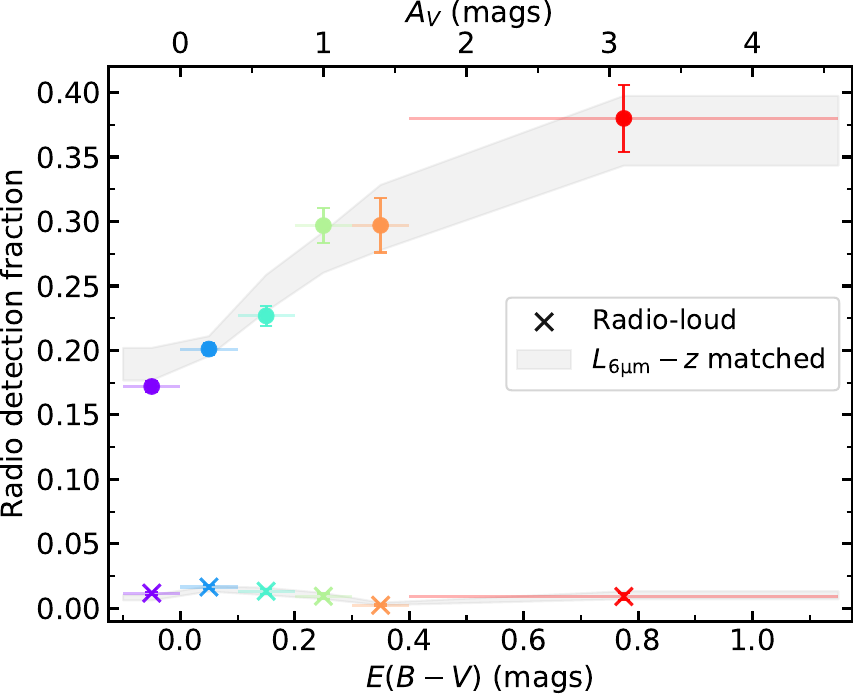}
    \caption{LDR2 radio-detection fraction and radio-loud ($\mathcal{R}\geq-4.5$) fraction for the combined QSO sample in bins of $E(B-V)$ (the fraction is calculated relative to the number of combined QSOs per $E(B-V)$ bin), calculated using a PL extinction curve (see Section~\ref{sec:dust_desi}). The grey shaded region displays the $L_{\rm 6\upmu m}$--$z$ matched bins. The error bars were calculated using the method described in \protect\cite{cam} and corresponds to 1$\sigma$ binomial uncertainties. There is a positive correlation between radio detection fraction and $E(B-V)$, even when accounting for any redshift or luminosity biases, demonstrating a striking relationship between the radio emission and the line-of-sight dust extinction towards QSOs. We find that the radio-loud fraction is $\lesssim$\,$2$ per~cent across all $E(B-V)$ bins, suggesting the correlation between radio-detection fraction and dust extinction is not due to large scale/high luminosity radio jets. There is a small decrease in the radio-loud fraction out to $E(B-V)$\,$\sim$\,0.4\,mag. The number of QSOs and associated radio detection and radio loud fractions can be found in Table~\ref{tab:radio_tab}.}
    \label{fig:av_radio}
\end{figure*}

\begin{table*}
    \centering
    \begin{tabular}{c|ccc|ccc}
        \hline 
        \hline
         \multirow{2}{*}{$E(B-V)$} & \multicolumn{3}{c|}{Combined QSO sample (all)} & \multicolumn{3}{c}{Combined QSO sample ($L_{\rm6\,\upmu m}$--$z$ matched)} \\
     & \# QSOs & Radio-detection frac. (\#) & Radio-loud frac. (\#) & \# QSOs & Radio-detection frac. (\#) & Radio-loud frac. (\#) \\
         \hline
         -0.1--0.0 & 7412 & 17.2\% (1275) & 1.2\% (87) & 861 & 18.9\% (163) & 0.8\% (7) \\
         0.0--0.1 & 7720 & 20.1\% (1552) & 1.6\% (127) & 2529 & 20.3\% (514) & 1.5\% (38) \\
         0.1--0.2 & 2725 & 22.7\% (618) & 1.3\% (36) & 843 & 24.4\% (206) & 1.3\% (11) \\
         0.2--0.3 & 1068 & 29.7\% (317) & 0.9\% (10) & 739 & 27.6\% (204) & 0.9\% (7) \\
         0.3--0.4 & 431 & 29.7\% (128) & 0.2\% (1) & 297 & 30.3\% (90) & 0.3\% (1) \\
         0.4--1.1 & 329 & 38.0\% (125) & 0.9\% (3) & 297 & 37.0\% (110) & 1.0\% (3) \\
         \hline
         \hline
    \end{tabular}
    \caption{Table displaying the number of QSOs, radio-detection fraction, and radio-loud ($\mathcal{R}\geq-4.5$) fraction in each $E(B-V)$ bin from Fig.~\ref{fig:av_radio}, for the full combined QSO sample and the $L_{\rm 6\upmu m}$--$z$ matched sample.}
    \label{tab:radio_tab}
\end{table*}

In Fig.~\ref{fig:av_radio} we show the LDR2 radio detection fraction for the combined QSO sample in bins of $E(B-V)$. On the basis of this figure it is clear that there is a strong positive correlation between the line-of-sight dust extinction and the radio detection fraction for DESI QSOs. This is consistent with our previous work in which we found that red QSOs in SDSS, selected by their $g-i$ optical colour, have a higher radio detection fraction compared to typical blue QSOs \citep{klindt,rosario,fawcett20,fawcett21}. To test that there are no redshift or luminosity biases that are driving this result, we matched the QSOs within each $E(B-V)$ bin in redshift and $L_{\rm 6\,\upmu m}$ (using a tolerance of 0.05 and 0.2\,dex for redshift and $L_{\rm 6\,\upmu m}$, respectively), using a 2:1 match ratio for the bluest bins to maximize source statistics. We show the LDR2 detection fraction for each matched bin in Fig.~\ref{fig:av_radio} by the grey shaded region; although limited on source statistics, a strong positive trend is clearly still present, even when accounting for redshift and luminosity biases (see Appendix~\ref{sec:test} for additional tests). The number of QSOs and associated radio detection fractions for the full and $L_{\rm 6\,\upmu m}$--redshift matched samples can be found in Table~\ref{tab:radio_tab}. It is also worth noting that no radio pre-selection is incorporated in the DESI QSO target selection, in comparison to SDSS where some QSOs are selected based on a FIRST radio detection (our previous work addressed this pre-selection; see Section~2.2.1 in \citealt{klindt}). This result therefore corroborates our previous SDSS red QSO results, but utilizing a cleaner, less biased sample and expanding to QSOs with even higher line-of-sight dust extinction (SDSS red QSOs span the range $0.03$\,$\lesssim$\,$E(B-V)$\,$\lesssim$\,0.3; \citealt{richards,klindt,fawcett20,calistro}).

\begin{figure*}
    \centering
    \includegraphics[width=0.97\textwidth]{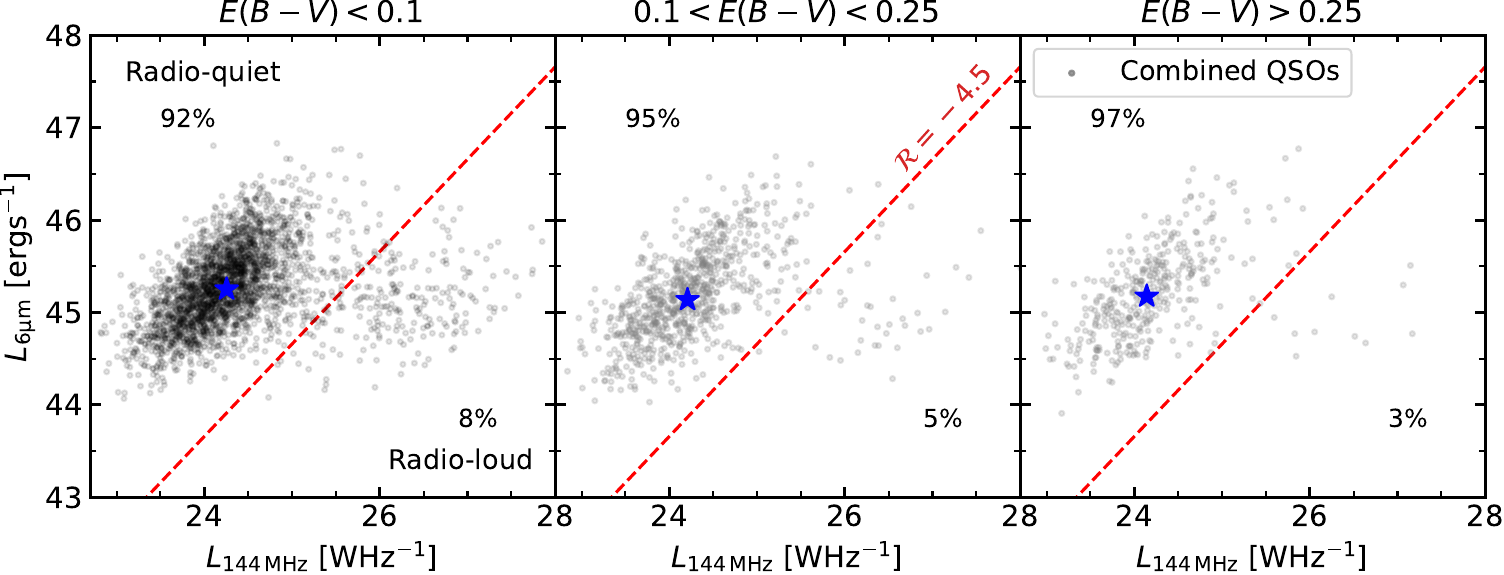}
    \caption{$L_{\rm 6\,\upmu m}$ versus $L_{\rm 144\,MHz}$ for the radio-detected combined QSO sample (see Section~\ref{sec:dust_desi_result}) in three bins of $E(B-V)$. The division between radio-loud and radio-quiet sources is displayed by the dashed red line ($\mathcal{R}=-4.5$; see Section~\ref{sec:radio_data}); $\sim$\,93 per~cent of the radio-detected combined QSO sample are radio-quiet (this increases to $\sim$\,99 per~cent when taking the radio-undetected QSOs into account, which are radio-quiet based on a $5\sigma$ LDR2 upper limit). The blue stars indicate the median value in each $E(B-V)$ bin; there is no significant change in median radio-loudness across the $E(B-V)$ bins for the combined QSO sample. However, we do find a significant change in the fraction of radio-detected QSOs that are radio-loud ($\mathcal{R}\geq-4.5$) across the $E(B-V)$ bins (indicated on the plot), with fewer radio-loud QSOs at higher $E(B-V)$ values; we further investigate this result in Section~\ref{sec:origin_dis}.}
    \label{fig:L6_Lrad}
\end{figure*}

To give some basic insight on the radio properties, in Fig.~\ref{fig:L6_Lrad} we show the $L_{\rm 6\,\upmu m}$ versus $L_{\rm 144\,MHz}$ distribution for the radio-detected combined QSO sample in three bins of $E(B-V)$, with the radio-loud/radio-quiet threshold displayed by the dashed red line (see Section~\ref{sec:radio_data}); we find that 93 per~cent of the radio-detected QSOs in the combined sample are radio-quiet. Assuming a $5\sigma$ upper limit for the LDR2 undetected QSOs (415\,$\upmu$Jy) we estimate a radio-quiet fraction for the overall combined QSO sample of $\sim$\,99 per~cent. Therefore, the trend in radio detection fraction found in Fig.~\ref{fig:av_radio} is driven by radio-quiet QSOs, consistent with our previous work. We do not find a significant difference in the median radio-loudness across increasing $E(B-V)$ bins ($\mathcal{R}$\,$\sim$\,$-5.9$ for each $E(B-V)$ bin). However, we do find a decrease in the number of radio-loud QSOs with increasing $E(B-V)$ ($\sim$\,$8$, 5, and 3 per~cent with increasing $E(B-V)$ bin); we explore this result more in Section~\ref{sec:origin_dis}.

\section{Discussion}\label{sec:dust_desi_rad_dis}
We have presented the first results from our ongoing secondary target program in DESI that uses an optical and MIR colour selection to expand the optical colour space probed by the nominal DESI QSO selection. Our program complements the QSOs from the nominal DESI QSO selections, which are predominantly blue, and pushes to higher extinctions than those selected by both the SDSS and DESI QSO samples. Combining the QSOs from our secondary target program and the nominal DESI QSOs, the key result in this paper is the striking positive relationship between amount of line-of-sight dust extinction and the radio detection fraction for QSOs (Fig.~\ref{fig:av_radio}). This clearly demonstrates that the presence of dust (and/or gas, i.e., opacity) is an important factor for the production of radio emission in radio-quiet QSOs. In the following sections we discuss how these results complement and extend the previous red QSO studies, including the overlap between our sample and Extremely Red Quasars (ERQs; e.g., \citealt{hamann}; Section~\ref{sec:red_dis}), and explore the origin of the radio emission in radio-quiet dusty QSOs (Section~\ref{sec:origin_dis}). 

\subsection{A comparison to previous red and obscured QSO studies}\label{sec:red_dis}

\begin{table*}
    \centering
    \begin{tabular}{c|cccc}
        \hline 
        \hline
         Sample name & Key selection & $E(B-V)$ & \# sources & Reference  \\
         &&[mag]&\\
         \hline
         \multirow{2}{*}{DESI combined QSOs} & \multirow{2}{*}{DESI QSO selections \& \cite{mateos} $W123$}  & 0.1--1.1 & 4553 & This work \\
          & & 0.25--1.1 & 1177 & This work \\
         \multirow{2}{*}{SDSS red QSOs} & DR7 $(g-i)$ 90th percentile & \multirow{2}{*}{$\lesssim0.2$} & 4899 & \cite{klindt} \\
         & DR14 $(g-i)$ 90th percentile & & 21\,800 & \cite{fawcett20} \\
         \textit{WISE} red QSOs & $W1-W2$\,$>$\,0.7 \& $W2-W3$\,$>$\,$2$ \& $W3-W4$\,$>$\,$1.9$ \& $E(B-V)$\,$>$\,$0.25$ & 0.25--1.2 & 21 & \cite{glik18} \\
         W2M red QSOs & $0.5$\,$<$\,$W1-W2$\,$<2$ \& $2$\,$<$\,$W2-W3$\,$<$\,$4.5$ \& $E(B-V)$\,$>$\,$0.25$ & 0.25--1.2 & 63 & \cite{glikman22} \\
         HRQs & $J-K$\,$>$\,$1.6$ \& $W1-W2$\,$>$\,0.85 & 0.5--1.5 & 38 &  \cite{ban} \\
         ERQs & $i-W3$\,$>4.6$ & $\lesssim0.7$ & 97 & \cite{hamann} \\
         Obscured & \multirow{2}{*}{SED fitting} & $\gtrsim2$ & 355 & \cite{andonie} \\
         IR red QSOs & & 0.05--1 & 223 & \cite{andonie} \\
         \hline
         \hline
    \end{tabular}
    \caption{Comparison of previous red QSO selections and their corresponding range in $E(B-V)$: the QSOs explored in this study, SDSS red QSOs (\citealt{klindt,fawcett20,rosario,calistro}; \citetalias{fawcett22}), IR-selected red QSOs \citep{glik18}, WISE-2MASS (W2M) red QSOs \citep{glikman22}, Heavily Reddened QSOs (HRQs) \citep{ban}, Extremely Red Quasars (ERQs) \citep{ross,hamann}, obscured QSOs and unobscured IR-selected red QSOs \citep{andonie}. For a more comprehensive comparison of red and obscured QSOs, see Table 2.1 within \protect\cite{klindt_thesis}.}
    \label{tab:red_comp_tab}
\end{table*}

\subsubsection{SDSS red QSOs}

\begin{figure*}
    \centering
    \includegraphics[width=0.97\textwidth]{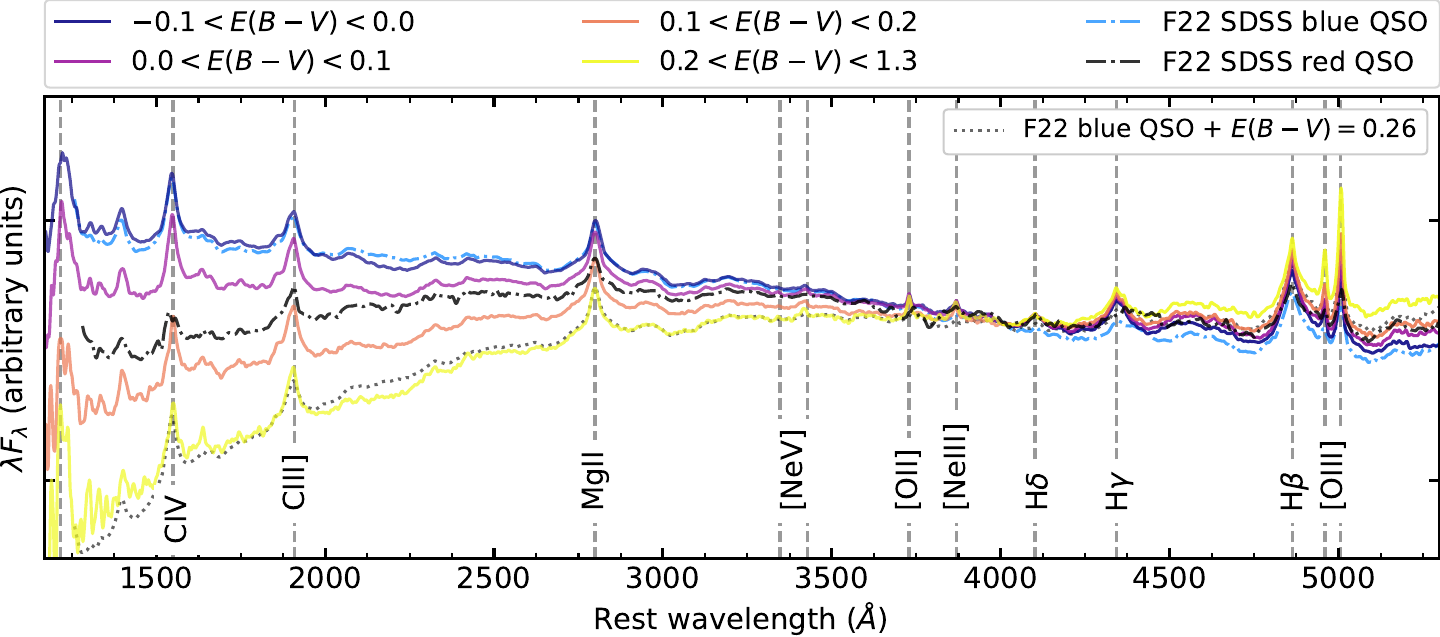}
    \caption{Comparison of composite spectra, normalized at 4000\,\AA, for the combined DESI QSO sample in luminosity--redshift matched bins of $E(B-V)$ (solid lines) to the \textit{X-shooter} red and blue composites from \citetalias{fawcett22} (dot-dashed black and blue lines, respectively). The \citetalias{fawcett22} blue composite with the average dust extinction for the reddest DESI bin applied is displayed by the dotted line; this agrees very well with the reddest DESI composite. We find that the DESI composites appear to be more dust reddened with increasing $E(B-V)$, and that the reddest composite has over twice the amount of average dust extinction as that of the \citetalias{fawcett22} red composite. This again demonstrates that our DESI SP will obtain a statistically significant sample of heavily reddened QSOs that are not observed by SDSS.}
    \label{fig:comp_av}
\end{figure*}

In our previous work exploring SDSS selected red and blue QSOs, we found that the majority of the red QSOs were red due to modest amounts of dust extinction ($E(B-V)\lesssim0.2$\,mag; \citealt{calistro}; \citetalias{fawcett22}). For example, utilizing VLT/\textit{X-shooter} data, \citetalias{fawcett22} found that the red QSO spectra displayed the characteristic frequency dependent spectral turnover at UV--optical wavelengths, indicating they were red due to line-of-sight dust extinction. To illustrate the improvements in the selection of dust-reddened QSOs by our DESI program, in Fig.~\ref{fig:comp_av} we show a comparison between the \textit{X-shooter} red and blue composites from \citetalias{fawcett22} and our combined DESI sample, in similar luminosity--redshift matched bins of measured $E(B-V)$ as in Fig.~\ref{fig:av_radio} (see Section~\ref{sec:dust_desi}). Firstly, we confirm that the DESI composites appear more dust reddened (i.e., a stronger UV spectral turnover) with increasing $E(B-V)$, providing confidence in our dust extinction fitting method. We also find that the reddest DESI composite has over $\sim$\,0.15\,mag more dust extinction on average compared to the \citetalias{fawcett22} SDSS red composite, again demonstrating that with DESI we can obtain a statistically significant sample of luminous heavily reddened QSOs that are missed by the optically shallow SDSS survey.

We have also previously found that SDSS red QSOs display enhanced radio emission compared to typical blue QSOs (\citealt{klindt,rosario,fawcett20,fawcett21}). Similarly, \cite{glikman22} used a \textit{WISE}--2MASS colour selection to define a sample of red QSOs with dust extinctions $E(B-V)$\,$>$\,$0.25$\,mag ($A_V$\,$\gtrsim$\,$0.8$\,mag; by this definition, the SDSS QSOs defined as ``red'' in our previous work would be classified as ``blue'' in the \cite{glikman22} sample) and found a higher radio detection fraction for their more heavily reddened red QSOs compared to their blue sample (see Figure~18 therein), using both FIRST and VLASS. These results are in agreement with Fig.~\ref{fig:av_radio}, where we have now extended to more extreme red QSOs, which suggests that the radio enhancement in red QSOs is linked to a higher amount of line-of-sight dust extinction. Table~\ref{tab:red_comp_tab} displays the number of combined QSOs with SDSS-like reddening ($E(B-V)$\,$>$\,$0.1$\,mag) and the number that satisfy the \cite{glikman22} red QSO cut ($E(B-V)$\,$>$\,$0.25$\,mag), resulting in 4533 and 1177 red QSOs, respectively. This demonstrates that the DESI survey will obtain a statistically significant sample of heavily reddened QSOs.

\subsubsection{Obscured QSOs}
Extending our previous red QSO work to fully obscured QSOs (e.g., Compton thick), \cite{andonie} compared the VLA 1.4 and 3\,GHz radio properties of a sample of obscured and unobscured QSOs in the COSMOS field (by their definition, even our extreme DESI red QSOs are unobscured) and found that the obscured QSOs have higher radio detection fractions compared to unobscured QSOs (see Figure~14 therein). However, interestingly they found that the obscured QSOs have consistent radio detection fractions to red QSOs. This could suggest that either 1) there is a ``saturation'' point in Fig.~\ref{fig:av_radio}, whereby increasing the amount of line-of-sight dust extinction towards a QSO will \textit{not} result in a higher radio detection fraction, 2) the origin of the radio emission is different in obscured and unobscured QSOs, or 3) the differences in the radio detection fraction are more modest at the higher sensitivity radio data in COSMOS. Utilizing the increased source statistics in future DESI data releases we can populate the highest $E(B-V)$ end of Fig.~\ref{fig:av_radio}, testing whether the radio detection fraction increases or does indeed start to plateau, as would be the case in the former scenario; by the end of the five year survey we expect to add additional bins at $E(B-V)>0.4$ in Fig.~\ref{fig:av_radio}.

\subsubsection{Extremely Red QSOs}
Since our DESI study pushes to more reddened QSOs than the previous SDSS red QSO studies, it is interesting to investigate the overlap between our DESI red QSOs and the population of Extremely Red Quasars (ERQs; see Table~\ref{tab:red_comp_tab} for a comparison of the red QSO selections discussed in this section), which are selected based on their optical--MIR colours, as compared to the pure optical or NIR selection approach often used to select red QSOs \citep{ross,hamann}. ERQs have been found to host some of the most powerful outflows identified in QSOs (e.g., \citealt{zakamska_2016,perrotta,lau}), and therefore have also been suggested as a candidate transitional population in the evolution of QSOs (see discussion in Section~\ref{sec:origin_dis}). However, due to the depth of SDSS, previous studies have only analyzed a small sample of the most luminous ERQs. In order to assess the capability of DESI (in particular, our SP) at selecting ERQs, we first defined an ERQ selection based on the optical--MIR colours of the SDSS ERQ sample from \cite{hamann}. \cite{hamann} defined an ERQ to have redshifts\,$>2.0$, $W3$ SNR\,$>3$, and $i-W3\geq4.6$\,mag. For our ERQ classification, we based our selection on the full \cite{hamann} ERQ sample that meet the $i-W3\geq4.6$\,mag selection; 141 objects. Since the DR9 photometry does not include an $i$-band magnitude, we defined our DESI ERQ selection based on the $z-W3$ colours, qualitatively consistent with the ERQ selection of \cite{ross} and \cite{hamann}; we classify a QSO as an ERQ if it has an optical--MIR colour that satisfies $z-W3>3.9$\,mag. This resulted in $\sim$\,$39$\% (714/1852) of our SP QSO sample classified as ERQs, in comparison to only $\sim$\,$7$\% (6167/88\,101) for the nominal QSO sample, demonstrating that our SP is also efficient at selecting ERQs; in future data releases we will build up a statistically significant sample of these rare systems. Our future papers will explore the dust extinction, emission-line, and radio properties of DESI ERQs and how they compare to the general red QSO population.

\subsubsection{Mg\,II Absorption Line Systems}\label{sec:mgII_dis}
A number of studies have noted that a larger fraction of reddened QSOs host narrow absorption lines compared to typical blue QSOs \citep{richards,Chen_2020,fawcett22}. The strongest and most distinctive of these absorption lines is the Mg\,{\sc ii}$\lambda\lambda 2796,2803$ doublet, with a laboratory separation between the two lines of 7.1772\,\AA~\citep{pickering}. Mg\,{\sc ii} absorption that is close to the Mg\,{\sc ii} emission line (i.e., where $z_{\rm abs}$\,$\sim$\,$z_{\rm em}$) is thought to be associated with the host galaxy of the QSO, whereas blueshifted absorption (i.e., where $z_{\rm abs}$\,$<<$\,$z_{\rm em}$) is thought to be due to an intervening system. Previous studies have found that QSOs that host associated absorption lines (AALs) are, on average, redder, with stronger [O\,{\sc ii}] emission, and are more likely to be radio detected, compared to both QSOs with intervening absorption lines and QSOs with no absorbers \citep{vanden_2008,shen_mernard,khare_2014,Chen_2020}. The strength of the Mg\,{\sc ii} absorption in systems with AALs has also been found to correlate with the dust extinction and radio properties, suggesting an intrinsic nature for the AALs \citep{vanden_2008,Chen_2020}. Despite the similarities in radio properties between QSOs with AALs and red QSOs, the measured dust extinctions are far more modest in the AAL QSOs, with an $E(B-V)$\,$\sim$\,$0.03$--0.06\,mag \citep{Chen_2020}. Furthermore, \cite{shen_mernard} suggested that AAL QSOs may be a transitional population between heavily reddened QSOs and typical blue QSOs. In order to assess whether AAL QSOs are driving the relationship between reddening and the radio detection fraction found in Fig.~\ref{fig:av_radio} and our previous SDSS work \citep{klindt,fawcett20,rosario}, we explored the properties of QSOs with Mg\,{\sc ii} absorbers in our DESI combined QSO sample, utilizing the catalogue from \cite{napolitano}.

\cite{napolitano} used an autonomous spectral pipeline to detect associated or intervening Mg\,{\sc ii} absorption through an initial line-fitting process and a Markov Chain Monte Carlo (MCMC) sampler. To create the Mg\,{\sc ii} absorber catalogue, they ran their detection pipeline on 83\,207 DESI QSOs at $0.3$\,$<$\,$z$\,$<$\,$2.5$ which are part of the DESI EDR. They found 20.1 per~cent (16\,707/83\,207) of the unique QSO spectra have Mg\,{\sc ii} absorbers, with a large number that contain multiple absorption features. 

\begin{figure}
    \centering
    \includegraphics[width=0.44\textwidth]{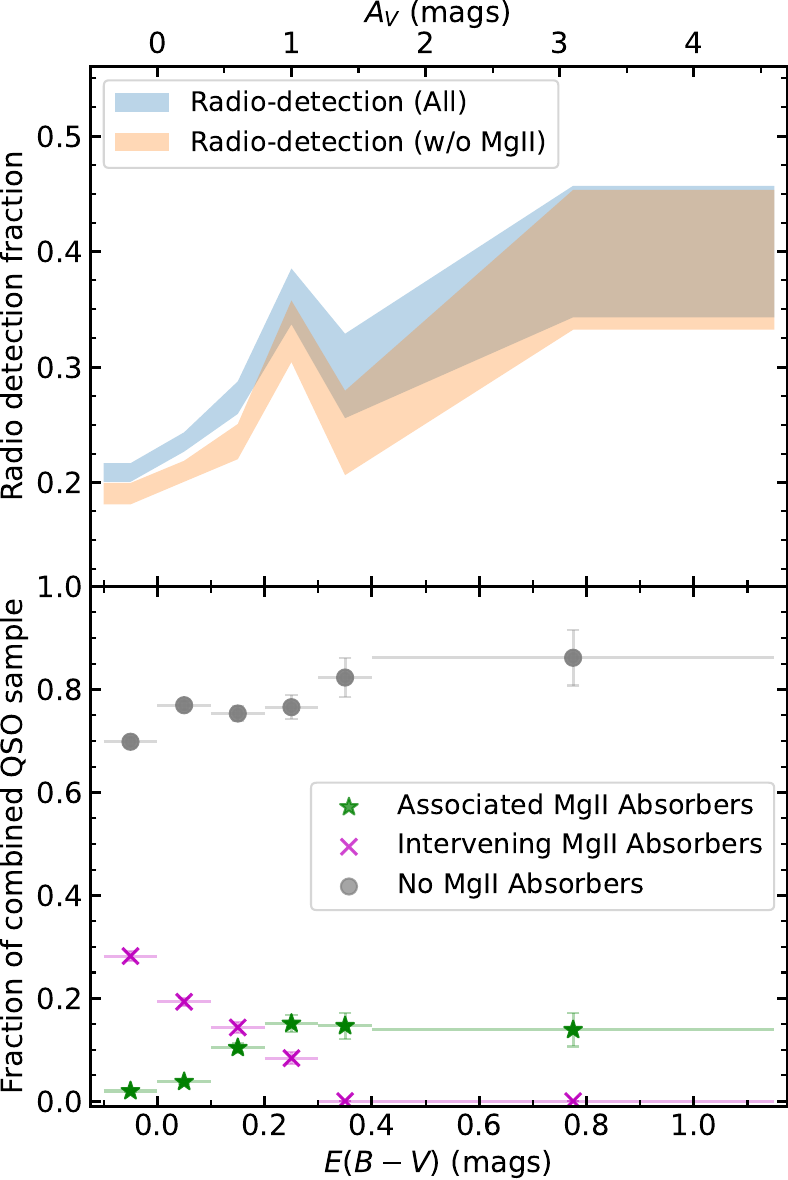}
    \caption{(Top) LDR2 radio-detection fraction including (blue) and excluding (orange) Mg\,{\sc ii} absorption systems, and (bottom) fraction of Mg\,{\sc ii} associated absorbers (green stars), intervening absorbers (magenta crosses), and no Mg\,{\sc ii} absorbers (grey circles) for the combined QSO sample, restricted to systems explored in \protect\cite{napolitano} (those included in the DESI EDR), as a function of $E(B-V)$. Excluding the systems with Mg\,{\sc ii} absorption present does not affect the overall trend between radio detection fraction and $E(B-V)$. The fraction of QSOs with associated Mg\,{\sc ii} absorbers increases with increasing $E(B-V)$, which is consistent with previous work that found QSOs with associated Mg\,{\sc ii} absorption tend to be redder and have higher radio detection fractions compared to both QSOs with intervening absorbers and those without any absorbers. The fraction of intervening Mg\,{\sc ii} absorbers decreases with increasing $E(B-V)$, which is likely due to the lower signal-to-noise in the reddest systems which may reduce the effectiveness of the automated spectral pipeline at detecting intervening absorbers.}
    \label{fig:radio_mgII}
\end{figure}

Restricting the combined QSO sample to those included in the EDR significantly reduces the sample to 11\,092 QSOs. Matching to the \cite{napolitano} Mg\,{\sc ii} catalogue resulted in 4924 absorption systems in 30.7 per~cent (3400/11\,092) unique QSO spectra. Of these, 16.4 per~cent (808/4924) are associated, defined by $-6000$\,$<$\,$v_{\rm off}$\,$<$\,$5000$\,km\,s$^{-1}$ ($v_{\rm off}=c\times\frac{z_{\rm MgII}-z_{\rm QSO}}{1+z_{\rm QSO}}$; \citealt{napolitano}), and the rest are intervening. Comparing the radio detection fractions, we found 24.3 per~cent, 13.4 per~cent, and 13.4 per~cent of the AAL QSOs, intervening absorber QSOs, and QSOs without any absorption systems have LDR2 radio detections, respectively. We also found the AAL QSOs are considerably redder than the intervening and no absorber systems, with a median $E(B-V)$\,$\sim$\,$0.12$, 0.00, and 0.03\,mag for the QSOs with associated, intervening, and no absorbers, respectively (see also Napolitano et~al. \textit{in prep}). Since the median $E(B-V)$ associated with the absorption feature has been found to be $\lesssim$\,$0.03$\,mag, this suggests that AALs are more likely to be associated with red QSOs. This is further shown in Fig.~\ref{fig:radio_mgII}, which displays the fraction of associated and intervening Mg\,{\sc ii} absorbers as a function of $E(B-V)$; a larger fraction of associated absorbers are found in QSOs with higher amounts of dust extinction, consistent with previous studies \citep{shen_mernard,Chen_2020,fawcett22}. This suggests the Mg\,{\sc ii} absorption is associated with the dust/gas that is also responsible for reddening the spectrum. We found that the fraction of QSOs with intervening absorbers decreases with increasing $E(B-V)$; by definition, intervening absorbers will fall at the bluer end of the spectrum (relative to Mg\,{\sc ii}), which will have a lower signal-to-noise for red QSOs due to dust extinction, and so this could reduce the effectiveness of the automated pipeline at detecting absorption features.

Fig.~\ref{fig:radio_mgII} also displays the radio-detection fraction for the sample with and without both the intervening and associated absorber systems. We find that removing the absorber systems has little affect on the overall trend with $E(B-V)$, which is consistent with Fig.~\ref{fig:av_radio} albeit with lower source statistics, suggesting that the amount of dust (and/or gas, i.e., opacity) in a QSO is the most important factor for determining whether a QSO is radio detected or not. However, since QSOs with more dust are more likely to host an AAL, QSO samples selected purely on the presence of Mg\,{\sc ii} AALs are likely to be redder and more radio-detected compared to QSOs without absorbers.

\subsection{Origin of the radio emission in dusty radio-quiet QSOs}\label{sec:origin_dis}

The strong relationship between the line-of-sight dust extinction and radio detection fraction displayed in Fig.~\ref{fig:av_radio} demonstrates an intrinsic link between dust and the production of radio emission in QSOs. However, it is currently unclear what is the dominant mechanism for the production of the radio emission. Calculating the fraction of radio-loud ($\mathcal{R}$\,$>$\,$-4.5$; see Section~\ref{sec:radio_data}) QSOs in the combined sample as a function of $E(B-V)$, we find that $\lesssim$\,$2$ per~cent of the QSOs (both radio-detected and undetected) are radio-loud across all bins (see Fig.~\ref{fig:av_radio} and Table~\ref{tab:radio_tab}). This demonstrates that radio-loud QSOs, typically hosting a large-scale radio jet, are not driving the correlation between dust extinction and radio-detection fraction, in agreement with our previous red QSO studies \citep{klindt,fawcett20,rosario}. 

The low frequency radio emission probed by LOFAR is dominated by synchrotron emission, which may be produced by star formation, low-powered/frustrated jets, and/or winds causing shocks in a dusty circumnuclear/ISM region (for more details on the production of radio emission in radio-quiet QSOs, see review by \citealt{pan}). To perform a detailed analysis into the origin of the radio emission, multi-frequency and/or higher spatial resolution radio observations are required; our future radio study utilizing radio data from the upgraded Giant Meter Radio Telescope (uGMRT; PI: V. Fawcett), in addition to high spatial resolution e-MERLIN data \citep{rosario_21}, aims to do just this. In the following section we discuss some of the potential scenarios that could be driving the connection between the line-of-sight dust extinction and radio detection fraction in DESI QSOs.

In our previous work exploring the properties of SDSS red QSOs, we found no differences in the star-formation properties between the red and blue QSOs. This was based on deep IR data from the COSMOS region \citep{fawcett20} and composite SEDs \citep{calistro}. We therefore concluded that the radio emission in SDSS red QSOs was likely driven by AGN processes such as winds or jets. However, given that DESI pushes to redder systems, it is unclear whether this conclusion still holds for our combined QSO sample. In addition to comparing obscured and unobscured QSOs in COSMOS, \cite{andonie} also compared the star-formation properties between a blue ($E(B-V)$\,$<$\,$0.05$\,mag) and red ($0.05$\,$<$\,$E(B-V)$\,$<$\,$1.0$\,mag) unobscured sub-sample. They found no difference in the star-formation properties between the two samples, despite finding a significantly higher radio-detection fraction for the red QSOs compared to the blue QSOs. Given that the red and blue QSOs from \cite{andonie} cover a similar $E(B-V)$ range to Fig.~\ref{fig:av_radio}, this suggests that star formation is not a strong driver of the relation between dust extinction and radio detection fraction. Furthermore, Calistro~Rivera et~al. (\textit{in prep}) perform comprehensive SED fitting of red and blue QSOs in the LOFAR deep fields and find no difference in the star-formation properties. However, individual star formation measurements are required to robustly demonstrate this for our sample. Calculating the $L_{\rm 144\,MHz}$ from star-forming galaxies with a star-formation rate of 100\,M$_{\rm \odot}$yr$^{-1}$ (equivalent to local ultraluminous infrared galaxies; ULIRGs), based on the FIR-radio correlation from \cite{calistro_17}, we find that only $\sim$\,16 per~cent of the radio-detected QSOs may have a substantial contribution from star formation to the radio emission, $\sim$\,83 per~cent of which are at $z$\,$<$\,$1$. Therefore, the most likely mechanism that links the presence of dust in the QSOs to the radio emission is either frustrated jets and/or winds causing shocks in a dusty circumnuclear/ISM region. 

Recent hydrodynamic simulations have predicted that radiation pressure on dust is an effective way to launch outflows (e.g., \citealt{costa}). The optical and UV radiation from the accretion disc is absorbed and re-emitted by the dust as IR radiation, transferring momentum to the ambient gas \citep{ishibashi,thompson}. These AGN-driven outflows can have high velocities ($\sim$\,$1000$--3000\,km\,s$^{-1}$), extend up to tens of kpc \citep{cicone}, and are likely launched along the polar axis (e.g., \citealt{hoenig}). When the outflow interacts with the surrounding dense ISM, they can cause shocks which accelerate electrons, producing radio emission \citep{Zubovas_2012,faucher,nims}. Dusty winds are therefore a promising mechanism for the production of radio emission in radio-quiet red QSOs. This is supported by a number of studies that have observed powerful outflows in red QSOs \citep{Urrutia_2009,hwang,perrotta,calistro,vayner,stacey}, although it is not always clear whether this is a result of luminosity biases (e.g., \citealt{villar}; \citetalias{fawcett22}). 

On the other hand, low-powered jet-ISM interactions have also been found to be important in radio-quiet QSOs \citep{jarvis,girdhar,Murthy2022,audibert}. For example, jet-ISM simulations have found that an initially collimated radio jet can get spread into lower powered jet filaments, which can potentially drive shocks in the ISM \citep{wagner,bicknell_18}. A significant fraction of the jet's energy (up to $30$ per~cent; \citealt{wagner}) can be imparted to the surrounding gas clouds, driving outflows. This is in agreement with observational studies that find correlations between the strength and/or spatial location of the ionized outflow with a young or weak radio jet \citep{cresci,cresci_23,morganti,molyneux_2019,venturi}. The connection between these low-powered radio jets and the expanding ionized gas bubbles have been shown to result in both negative and positive feedback in the host galaxy \citep{gaibler,jarvis,girdhar}, demonstrating the importance for studying the radio-ISM interactions in these radio-quiet systems. 

With our current data it is difficult to distinguish between the jet versus wind scenarios (or whether it is some combination of both). However, in either case the resulting effect is a jet/wind shocking the surrounding ISM. This will likely drive out the surrounding gas/dust in a short-lived ``blow-out'' phase (e.g., \citealt{glik17}) that may eventually reveal a typical blue QSO \citep{hop6}. Therefore, the relationship between dust and radio emission in QSOs may be crucial to our understanding of QSO evolution.

In future DESI papers we will robustly compare the emission line and radio properties of a statistically significant sample of QSOs as a function of $E(B-V)$, in order to understand to what extent differences in the observed outflow properties and/or accretion state drive the striking relationship in Fig.~\ref{fig:av_radio}. In addition to our future DESI studies, we have obtained four-band radio data observed with the uGMRT (PI: V. Fawcett) for a sample of 20 red and 20 blue radio-detected QSOs with high spatial resolution e-MERLIN 1.4\,GHz data from \cite{rosario_21}. With these data we can construct sensitive radio SEDs over 0.12--3\,GHz (with the addition of VLASS), in order to search for spectral signatures of radio jets (e.g., \citealt{bicknell_18}). For example, if the QSOs are similar to compact steep spectrum (CSS) and gigahertz-peaked spectrum (GPS) sources (see review by \citealt{O_Dea}), which are thought to be jet dominated, then we can expect a spectral turnover in the radio SED around $\sim$\,$100$\,MHz and $\sim$\,$1$\,GHz (for CSS and GPS sources, respectively). Therefore, we can explore 1) whether the radio SEDs of red and blue QSOs are different, and 2) if jets are the dominant mechanism for producing radio emission in radio-quiet dusty QSOs. Furthermore, future Very Long Baseline Interferometry (VLBI) observations will be crucial to resolve and confirm potential pc-scale radio jet morphologies in these systems.

\section{Conclusions}
In this paper, we present the first $\sim$\,$8$ months of our DESI secondary target program that utilizes a MIR and optical selection to expand the DESI QSO colour space, in particular to reddened QSOs. Combining our SP QSO sample with the larger nominal DESI QSO sample, we have explored the dust extinction and radio properties. From our analyses we find that:

\begin{itemize}
    \item \textbf{Our secondary target program extends the nominal DESI QSO program to more extreme optical colours, the majority of which are red}: From visually inspecting the spectra of the 3038 objects from our secondary target program, we find that $\sim$\,$89$ per~cent have high quality spectral and redshift classifications. Of these high quality sources, 81 per~cent are QSOs (both Type 1 and Type 2), 13 per~cent are galaxies, and 6 per~cent are stars (Fig.~\ref{fig:spec_pie_VI}). Overall, our program provides $\sim$\,$32$ per~cent more QSOs that lie outside of the main colour space compared to the nominal QSO sample despite being $\sim$\,$2$ per~cent of the size (Fig.~\ref{fig:colour_colour2}), and also pushes to fainter optical magnitudes than the nominal QSO program (Fig.~\ref{fig:L6_red}). See Sections~\ref{sec:method_desi_paper}, \ref{sec:qso_sample} and \ref{sec:red_dis}. 
    \item \textbf{DESI is capable of detecting QSOs with up to $E(B-V)$\,$\sim$\,$1$\,mag of dust extinction:} Measuring the dust extinction by fitting the spectra of the combined QSOs with a dust-reddened blue QSO composite, we find up to $\sim$\,1 mag in $E(B-V)$ ($A_V$\,$\sim$\,4\,mag; Figs.~\ref{fig:av_hist_all}), comparable to NIR-selected heavily reddened QSOs. We also find that optical colour strongly correlates with dust extinction, demonstrating that optical colour is an effective way to select dust-reddened QSOs (Fig.~\ref{fig:gr_rz_colour}). Finally, splitting the combined QSO sample into $E(B-V)$ bins, we find that the reddest composite ($E(B-V)$\,$>$\,$0.2$\,mag) has over $\sim$\,0.15\,mag more dust extinction than the \cite{fawcett22} SDSS red QSO composite, again demonstrating that with DESI we can push to much higher amounts of dust extinction than in SDSS (Fig.~\ref{fig:comp_av}). 
    See Sections~\ref{sec:dust_desi_result} and \ref{sec:red_dis}.
    \item \textbf{There is a strong link between the radio detection fraction and amount of line-of-sight dust extinction in QSOs:} Utilizing the combined QSO sample to maximize the source statistics, we explored the 144\,MHz radio properties of the QSOs using the LoTSS DR2 (LDR2) dataset. Comparing the LDR2 radio detection fraction of the combined QSO sample as a function of $E(B-V)$ we find a strong positive correlation (Fig.~\ref{fig:av_radio}), even when accounting for any redshift or luminosity biases, demonstrating that the amount of line-of-sight dust extinction is closely linked to the radio emission in a QSO. We find that $\lesssim$\,$2$ per~cent of the combined QSO sample are radio-loud (Fig.~\ref{fig:L6_Lrad}), suggesting that large-scale radio-loud jets are not driving the relation between radio-detection fraction and dust extinction. Therefore, we suggest the origin of this predominantly radio-quiet emission is likely due to frustrated jets or outflows causing shocks in a dust/gas rich ISM, which may drive away the dust in a ``blow-out'' phase. See Sections~\ref{sec:radio_desi_result} and \ref{sec:origin_dis}. 
\end{itemize}

We have demonstrated that DESI will observe a statistically significant sample of dust-reddened QSOs, bridging the gap between the more modest SDSS red QSOs and the heavily reddened NIR-selected QSOs. From our results we have shown that red QSOs are likely undergoing a dusty blow-out phase, in which winds and/or jets clear out the surrounding dust and gas, eventually revealing a blue, typical QSO. Therefore, red QSOs are important objects to study in the context of AGN feedback and QSO evolution.
In future papers we will use our growing DESI sample to investigate whether there are any differences in the outflow properties of QSOs as a function of dust extinction and to explore statistically significant sub-samples of rare/exotic QSOs (e.g., Extremely Red Quasars and Low-ionization Broad Absorption Line QSOs).

\section{Acknowledgments}
We would like to thank the anonymous referee for their constructive comments that greatly helped improved the clarity and structure of the paper.

VAF and CMH acknowledge funding from an United Kingdom Research and Innovation grant (code: MR/V022830/1) and (VAF) previously a quota studentship through grant code ST/S505365/1 funded by the Science and Technology Facility Council. DMA and ACE acknowledges the Science and Technology Facilities Council (through grant code ST/P000541/1). AB supported by the U.S. Department of Energy, Office of Science, Office of High Energy Physics, under Award No. DESC0009959. DR acknowledges the support of STFC grant NU-012097. ADM was supported by the U.S. Department of Energy, Office of Science, Office of High Energy Physics, under Award Number DE-SC0019022. We would like to thank E. Verner for providing us with the \ion{Fe}{ii} template presented in \cite{Verner_2009}.

This work uses observations collected at the European Southern Observatory under ESO programme 0101.B-0739(A).

LOFAR data products were provided by the LOFAR Surveys Key Science project (LSKSP; https://lofar-surveys.org/) and were derived from observations with the International LOFAR Telescope (ILT). LOFAR (van Haarlem et al. 2013) is the Low Frequency Array designed and constructed by ASTRON. It has observing, data processing, and data storage facilities in several countries, which are owned by various parties (each with their own funding sources), and which are collectively operated by the ILT foundation under a joint scientific policy. The efforts of the LSKSP have benefited from funding from the European Research Council, NOVA, NWO, CNRS-INSU, the SURF Co-operative, the UK Science and Technology Funding Council and the Jülich Supercomputing Centre.

This material is based upon work supported by the U.S. Department of Energy (DOE), Office of Science, Office of High-Energy Physics, under Contract No. DE–AC02–05CH11231, and by the National Energy Research Scientific Computing Center, a DOE Office of Science User Facility under the same contract. Additional support for DESI was provided by the U.S. National Science Foundation (NSF), Division of Astronomical Sciences under Contract No. AST-0950945 to the NSF’s National Optical-Infrared Astronomy Research Laboratory; the Science and Technologies Facilities Council of the United Kingdom; the Gordon and Betty Moore Foundation; the Heising-Simons Foundation; the French Alternative Energies and Atomic Energy Commission (CEA); the National Council of Science and Technology of Mexico (CONACYT); the Ministry of Science and Innovation of Spain (MICINN), and by the DESI Member Institutions: \url{https://www.desi.lbl.gov/collaborating-institutions}. Any opinions, findings, and conclusions or recommendations expressed in this material are those of the author(s) and do not necessarily reflect the views of the U. S. National Science Foundation, the U. S. Department of Energy, or any of the listed funding agencies.

The authors are honored to be permitted to conduct scientific research on Iolkam Du’ag (Kitt Peak), a mountain with particular significance to the Tohono O’odham Nation.

\section{Data availability}

The SV data and spectra underlying this article are publicly available in the DESI Early Data Release \citep{sv}, with the remaining data and spectra obtained from the Main survey publicly available in a future data release (DESI collaboration et~al. in preparation 2024).

The radio and VLT/\textit{X-shooter} data used in this study are publicly available in online archives.

The data underlying the figures in this paper are available online: \url{https://doi.org/10.5281/zenodo.8147342}.

Electronic tables containing the DESI \texttt{TARGETID} for the full combined QSO and SP samples can be found online. Additional redshift and photometric information are provided for the sources included in the DESI Early Data Release \citep{sv}.

\bibliography{bib}
\bibliographystyle{mnras}

\appendix

\section{Testing the performance of the modified pipeline}\label{sec:appendix_MP}

\begin{table*}
    \centering
    \begin{tabu} to 0.6\textwidth{XXXXX|X}
        \hline 
        \hline
        &&\multicolumn{3}{c|}{VI} & \\
        & & QSO & GALAXY & STAR & \multicolumn{1}{c}{Total} \\
        \hline
        \multicolumn{1}{c}{\multirow{3}{*}{Modified pipeline}} & QSO & 1389 & 9 & 0 & \multicolumn{1}{c}{1398 (53\%)}\\
         & GALAXY & 753 & 336 & 4 &  \multicolumn{1}{c}{1093 (41\%)} \\
         & STAR & 12 & 0 & 155 & \multicolumn{1}{c}{167 (6\%)} \\
         \hline
         & Total & 2154 (81\%) & 345 (13\%) & 159 (6\%) & \multicolumn{1}{c}{2658} \\
         \hline
         \hline
    \end{tabu}
    \caption{Table displaying the number of high-quality objects (removing the 42 systems with two objects in the same spectrum; see Section~\ref{sec:method_VI}) that were classified as a QSO, GALAXY, or STAR in the modified pipeline compared to the VI classifications; we found that 765 of the visually identified QSOs are missed by the modified pipeline (classified as a GALAXY or STAR). Overall, the modified pipeline correctly identifies $\sim$\,71 per~cent of spectypes.}
    \label{tab:spec_tab_comp}
\end{table*}

Comparing the redshifts and spectral types (``spectypes'') for the DESI SP sample obtained from the VI to the modified pipeline, we found that the modified pipeline correctly identifies $\sim$\,$71$ per~cent of all spectypes (see Table.~\ref{tab:spec_tab_comp}). We also found that $>$\,$72$ per~cent of our sample are high-quality QSOs compared to $\sim$\,$47$ per~cent classified by the modified pipeline; this is predominantly due to the number of reddened QSOs in our sample (see Section~\ref{sec:sample_desi} and \citealt{VI1}) and highlights the necessity of VI to reliably classify red QSOs in our sample. 

\begin{figure}
    \centering
    \includegraphics[width=0.47\textwidth]{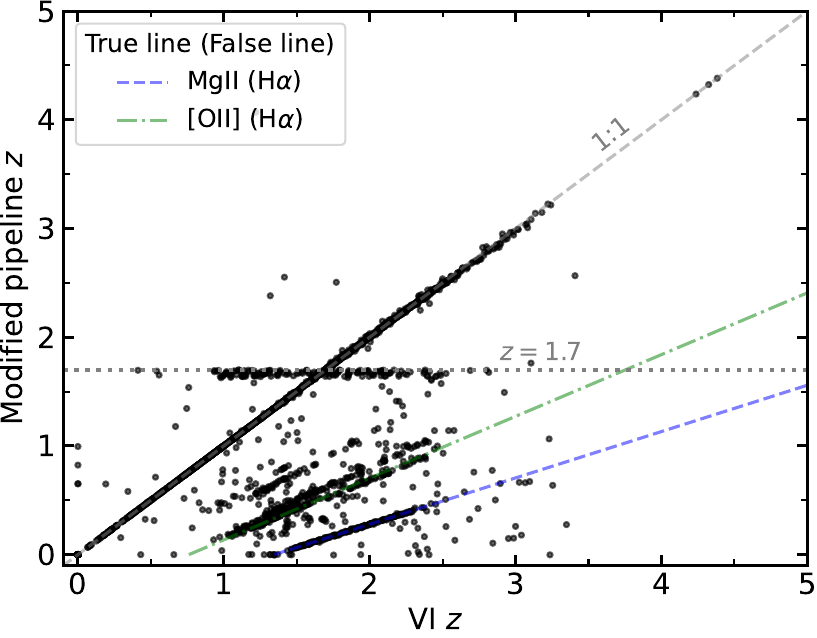}
    \caption{Redshifts obtained from the modified pipeline versus the redshifts obtained from VI (see Section~\ref{sec:method_VI}); we find that $\sim$\,66 per~cent of the high-quality objects have correct redshifts ($\Delta v\leq3000$\,kms$^{-1}$) as measured by the modified pipeline. The green and blue dashed lines highlight two redshift tracks caused by the modified pipeline incorrectly identifying the Mg\,{\sc ii} emission line as H\,$\alpha$ and [O\,{\sc ii}]$\uplambda$3727, respectively. The dashed grey line represents the 1:1 relation. The horizontal dotted grey line indicates the maximum redshift that the modified pipeline can assign to a galaxy ($z$\,$\sim$\,$1.7$; QSOs do not have a redshift restriction in the modified pipeline); the cluster of sources around this line therefore represents QSOs misidentified as galaxies.}
    \label{fig:red_tracks}
\end{figure}

A comparison between the resulting high-quality VI redshifts and those from the modified pipeline for our DESI SP sample is displayed in Fig.~\ref{fig:red_tracks}. We found $\sim$\,$66$ per~cent (1795/2700) of all high-quality objects have correct redshifts ($\Delta z=|z_{\rm VI}-z_{\rm RR}|/(1+z_{\rm VI})<0.01$, which corresponds to a $\Delta v\leq3000$\,kms$^{-1}$) as measured by the modified pipeline. In the SP QSO sample, $\sim$\,$58$ per~cent (1076/1852) have correct redshifts from the modified pipeline. Of the other $\sim$\,$42$ per~cent, $\sim$\,$78$ per~cent were also incorrectly classified as galaxies; these systems are all assigned a redshift $<$\,$1.7$ by the modified pipeline, due to the maximum redshift that the modified pipeline can assign to a galaxy. The fraction of correct redshifts in our SP QSO sample is a lot lower than that found in the VI of the nominal DESI QSOs ($\sim$\,$99$ per~cent); this is again predominantly due to the high number of dust-reddened QSOs in our sample, which are substantially more likely to be miss-classified \citep{VI1}.

\section{Population stacks}\label{sec:properties_stack}

\begin{figure*}
    \centering
    \includegraphics[width=0.95\textwidth]{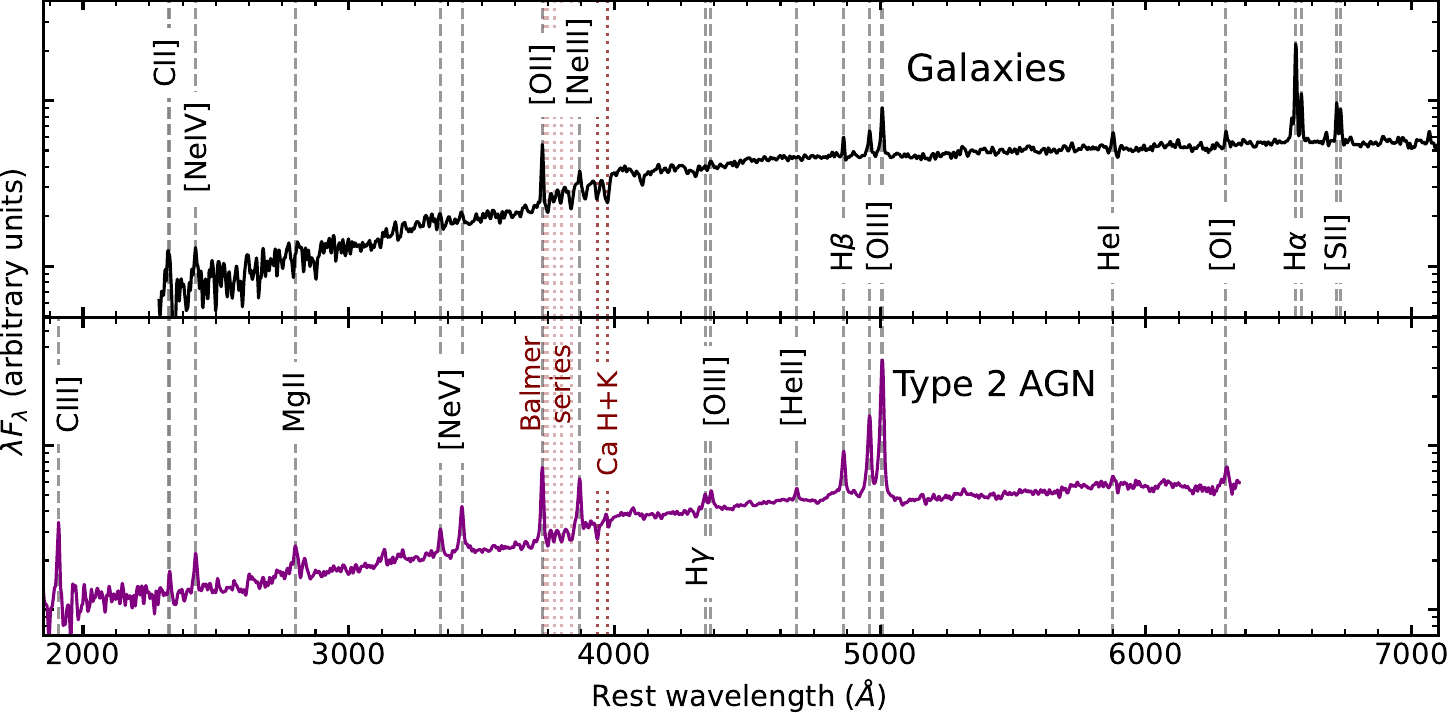}
    \caption{Composite of 358 and 306 high-quality VI-confirmed galaxies (top) and Type 2 AGN (bottom) in our DESI SP sample (see Section~\ref{sec:sample_desi}). The main emission and absorption features are indicated; the prominent [Ne\,{\sc v}]$\uplambda\uplambda$3346,3426 emission lines in the Type 2 AGN is strong evidence for AGN activity.}
    \label{fig:type2_comp}
\end{figure*}

Fig.~\ref{fig:type2_comp} displays the composite of the 358 and 306 high-quality galaxies and Type 2 AGN in our DESI SP sample (see Section~\ref{sec:sample_desi}), following the approach outlined in Section~\ref{sec:comp}. The Type 2 AGN display strong high ionization line emission lines, including [Ne\,{\sc v}]$\uplambda\uplambda$3346,3426, that implies the presence of hard radiation from an AGN \citep{mignoli}. In the galaxy composite, the emission lines are narrower compared to the Type 2 composite and the [Ne\,{\sc v}] lines are faint-to-non-existent, demonstrating little-to-no AGN activity in these objects. For a more detailed discussion of the galaxies in DESI, see \cite{bgs,bgs2,LRG2,lrg,ELG2,elg,lan}. The broader [O\,{\sc iii}] in the Type 2 composite compared to the galaxy composite implies that the Type 2 AGN are more likely to host powerful outflows; future work will investigate the emission line properties of our SP sample.

\begin{figure*}
    \centering
    \includegraphics[width=0.95\textwidth]{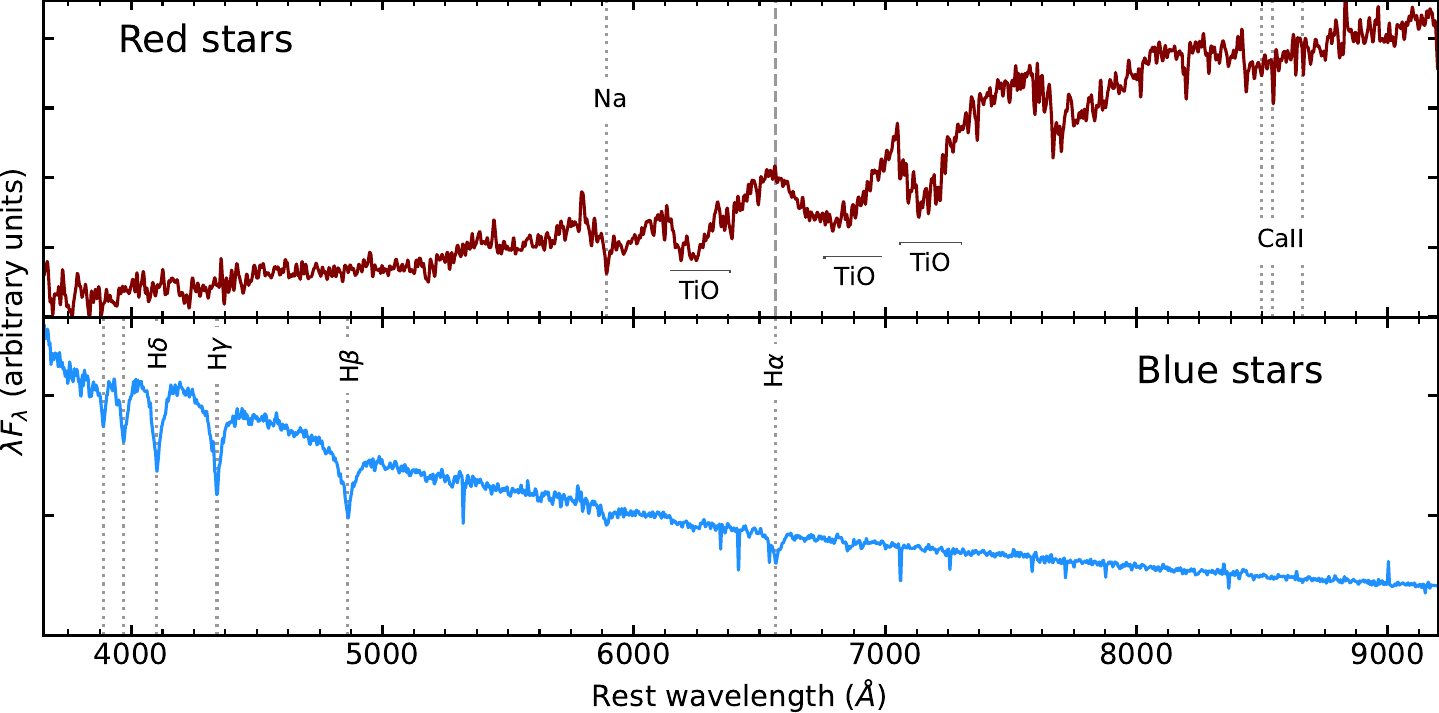}
    \caption{Composite of the 164 high-quality VI-confirmed stars in our DESI SP sample (see Section~\ref{sec:sample_desi}), split into red (top; $r-z$\,$>$\,$-0.4$, 142 stars) and blue (bottom; $r-z$\,$<$\,$-0.4$, 22 stars). The red stars show strong molecular features, indicative of red late-type stars (e.g., M-type), whereas the blue stars are mainly white dwarfs, as evidenced from the broad yet shallow Balmer absorption lines. The main emission and absorption features are indicated.}
    \label{fig:star_comp}
\end{figure*}

Fig.~\ref{fig:star_comp} displays the composite of the 164 high-quality stars in our DESI SP sample, split into ``red'' ($r-z$\,$>$\,$-0.4$) and ``blue'' ($r-z$\,$<$\,$-0.4$). The red stars show strong molecular features that are indicative of late-type cool stars (e.g., M-type), whereas the blue stars are mainly (DA) white dwarfs due to the broad and shallow Balmer features (see Figure~11 in \citealt{WD}).\footnote{For examples of different types of stellar spectra, see \url{http://classic.sdss.org/dr5/algorithms/spectemplates/}.} For a more detailed discussion of the stars in DESI, see \cite{stars,mws}.

\section{SDSS and DESI comparison}\label{sec:sdss_desi}

\begin{table}
    \centering
    \begin{tabular}{ccccc}
        \hline
        \hline
        & \multicolumn{2}{c}{Redshift} & \multicolumn{2}{c}{Spectype} \\
        SDSS ID & SDSS & DESI VI & SDSS & DESI VI \\
        \hline
        131729.28+545720.6 & 5.07 & 1.65 & QSO & QSO \\
        153420.23+413007.6 & 0.92 & 1.36 & QSO & QSO \\
        155633.78+351757.3\textsuperscript{\textdagger} & 1.49 & 3.24 & QSO & QSO \\
        155921.70+435700.6 & 0.66 & 1.36 & QSO & QSO \\
        160113.56+422637.7 & 0.15 & 1.72 & QSO & QSO \\
        163148.98+495933.4 & 4.53 & 1.45 & QSO & QSO \\
        170016.63+350353.5 & 4.50 & 1.39 & QSO & QSO \\
        \hline
        142843.33+120913.3 & 1.19 & 1.19 & QSO & GALAXY \\
        \hline
        143827.41+370737.3 & 1.01 & 1.01 & QSO & QSO (Type 2) \\
        130100.88+320727.4 & 0.51 & 0.51 & QSO & QSO (Type 2) \\
         \hline
         \hline
    \end{tabular}
    \caption{Table containing the SDSS ID, SDSS and DESI VI redshift, and SDSS and DESI VI spectral type for the ten QSOs which either had a redshift or spectral type disagreement between the SDSS DR16 catalogue and that from the VI. \textsuperscript{\textdagger}BALQSO.}
    \label{tab:sdss_miss}
\end{table}

Matching our SP sample to the SDSS DR16 QSO catalogue \citep{dr16} we found 117 matches, of which 110 are considered high-quality from the VI (107 are included in our SP QSO sample; see Section~\ref{sec:sample_desi}). For these 110 matches, we compared our VI redshifts and spectypes to that from the SDSS DR16 catalogue. Overall, seven sources have \textit{significantly} different redshifts in SDSS to those obtained by the DESI VI ($\Delta z=|z_{\rm VI}-z_{\rm SDSS}|/(1+z_{\rm VI})\geq0.05$; $\Delta v>15\,000$\,km\,s$^{-1}$), two sources were classified as Type 2 AGN in the VI and one additional source was classified as a galaxy in the VI, although all three had correct redshifts (see Table~\ref{tab:sdss_miss}). For these ten sources, we performed additional VI to confirm the redshift and spectype; the seven sources with redshift disagreements had clearly incorrect SDSS redshifts (including one extreme high redshift BALQSO; see Table~\ref{tab:sdss_miss}). The sources that were VI'd as a galaxy or Type 2 AGN have weak broad emission-line features.

\section{Testing the radio detection fraction plot}\label{sec:test}
\begin{figure}
    \centering
    \includegraphics[width=0.45\textwidth]{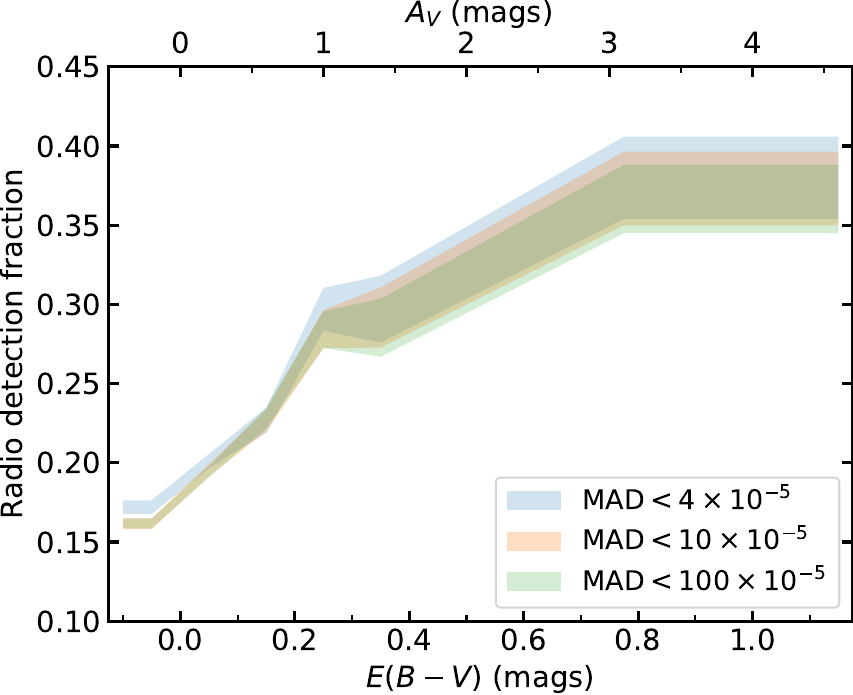}\vspace{0.8em}
    \includegraphics[width=0.45\textwidth]{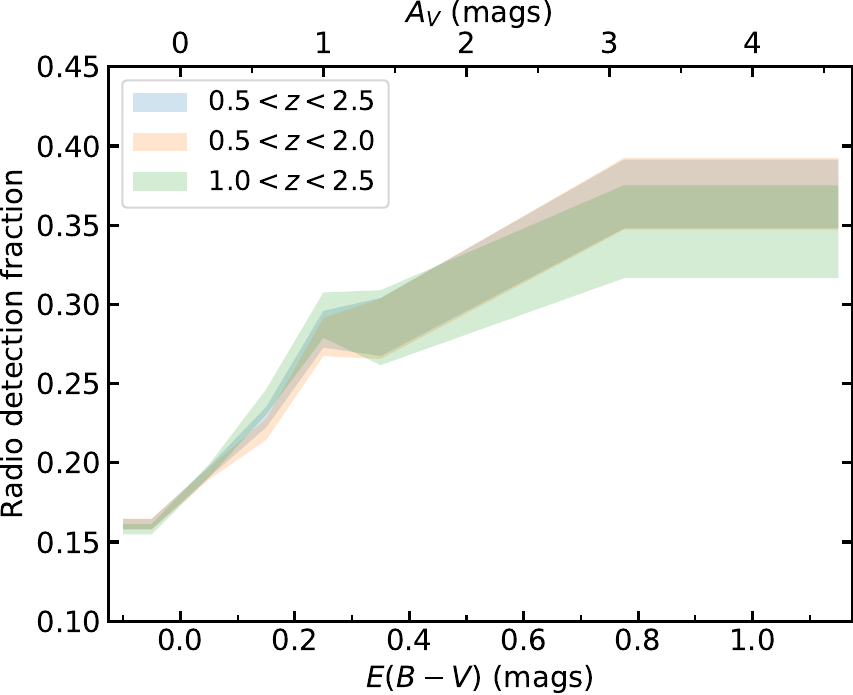}\vspace{0.8em}
    \includegraphics[width=0.45\textwidth]{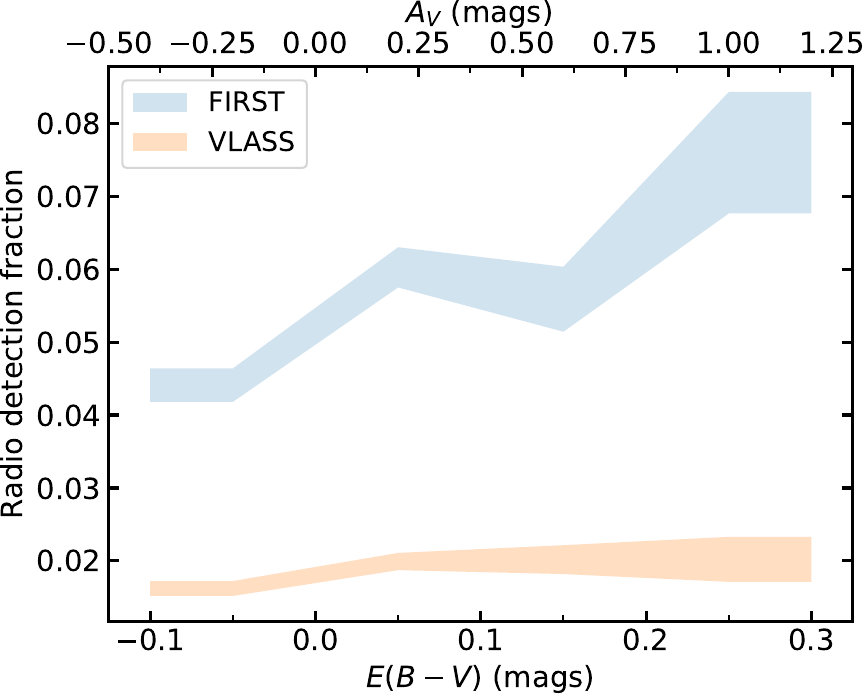}
    \caption{Similar to Fig.~\ref{fig:av_radio} with varying MAD cuts (top), varying redshift cuts (middle), and using both VLA FIRST and VLASS data (bottom). Overall, varying the MAD cut has little effect on the shape of the correlation. We find that applying a redshift cut of $0.5$\,$<$\,$z$\,$<$\,$2.0$ or $1.0$\,$<$\,$z$\,$<$\,$2.5$ does not effect the overall trend, although the source statistics are worse in the latter bin. Although limited by source statistics, the FIRST detection fraction also increases with increasing dust extinction (there are too few sources at $E(B-V)$\,$>$\,$0.3$\,mag to constrain the detection fraction). The VLASS detection fraction shows a slight increase, but this is not significant due to the limited source statistics. Examples of dust extinction fits with a range of MAD values can be found in the online supplementary material.}
    \label{fig:radio_test}
\end{figure}

The key result of this paper is the striking positive correlation between the amount of line-of-sight dust extinction and the radio detection fraction (Fig.~\ref{fig:av_radio}). To check the robustness of this result we performed three tests: 1) varying the MAD cut we imposed to define a ``good'' spectral fit, 2) changing the redshift cut, and 3) exploring the 1.4\,GHz FIRST and 3\,GHz VLASS detection fractions (see Fig.~\ref{fig:radio_test}). 

For the first test, we applied three different MAD cuts: $4\times10^{-5}$ (used for Fig.~\ref{fig:av_radio}), $7\times10^{-5}$, and $20\times10^{-5}$, which were satisfied by $\sim$\,58 per~cent (20\,031/34\,293), $\sim$\,85 per~cent (29\,001/34\,293), and $\sim$\,99 per~cent (33\,913/34\,293) of the combined QSOs, respectively. We find that relaxing the MAD cut slightly decreases the overall radio detection fraction, but does not change the shape of the trend; this demonstrates that the choice of MAD cut does not significantly affect our basic result. For the second test, we used a redshift cut of $0.5$\,$<$\,$z$\,$<$\,$2.0$ and $1.0$\,$<$\,$z$\,$<$\,$2.5$. The first range was chosen to exclude the highest redshift sources ($z$\,$>$\,$2.0$) that may not have such reliable fits due to the limited rest-frame spectral coverage of our blue QSO composite used in the fitting and the second range excludes the lowest redshift sources ($z$\,$<$\,$1.0$) that may still have contribution from the host galaxy (Section~\ref{sec:dust_desi}). We find that the overall trend is the same for both redshift ranges. 

\begin{table}
    \centering
    \begin{tabular}{cc}
         \hline \hline
         Radio survey & Radio detection fraction \\
         \hline
         LDR2 & 6549/34\,293 (19\%) \\
         FIRST & 1001/18\,816 (5\%)  \\
         VLASS &  631/34\,293 (2\%) \\
         \hline \hline
    \end{tabular}
    \caption{Radio detection fraction of the combined QSO table utilizing radio data from the LDR2, FIRST, and VLASS, adopting a matching radius of $5''$, $10''$, and $10''$, respectively. To calculate the FIRST radio detection fraction the combined QSO sample was restricted to the FIRST survey region, reducing the sample to 18\,816.}
    \label{tab:radio_det_tab}
\end{table}
For the last test, we matched the combined QSO sample to the 1.4\,GHz VLA FIRST and 3\,GHz VLASS\footnote{Note: for VLASS we utilized the ``Host ID'' table, which contains 700\,212 sources that have relatively simple morphology for which a host has been identified using a maximum likelihood ratio. This has been found to be more reliable than the larger ``Component'' table. The various tables can be found here: \url{https://cirada.ca/vlasscatalogueql0}} surveys, adopting a $10''$ matching radius for both surveys which resulted in 5 per~cent (1001/18\,816; the sample was reduced to the FIRST survey area) and 2 per~cent (631/34\,293) matches, respectively. This is consistent with the sensitivity of FIRST (0.15\,mJy) and VLASS (0.12\,mJy); assuming a canonical spectral index of $\alpha=0.7$, extrapolating the LDR2 flux values to 1.4 and 3\,GHz we expect 1079 and 882 QSOs to be detected in FIRST and VLASS, respectively. We find a positive trend in the FIRST radio detection fraction as a function of $E(B-V)$ out to $E(B-V)=0.3$\,mag, after which the source statistics are too limited to conclude anything meaningful (Fig.~\ref{fig:radio_test}). This is despite FIRST being $\sim$\,$9$ times less sensitive than LoTSS, again adding to the robustness of our result. The VLASS radio detection fraction displays a subtle increase with $E(B-V)$, but unfortunately the source statistics are too low to robustly conclude anything. Table~\ref{tab:radio_det_tab} displays the radio detection fractions of the combined QSO sample for LDR2, FIRST, and VLASS. 

\bsp	
\label{lastpage}
\end{document}